\documentclass[a4paper,11pt]{article}
\pdfoutput=1 

\usepackage{jheparxiv} 

\usepackage[T1]{fontenc} 

\usepackage{float}
\usepackage{tensor}
\usepackage{amsmath}
\usepackage{amssymb}
\usepackage{mathrsfs}
\usepackage{graphicx}
\usepackage{stmaryrd}
\usepackage{twistor}

\newcommand{\veps}{\varepsilon}
\newcommand{\msf}[1]{\mathsf{#1}}

\title{Supersymmetric celestial OPEs and soft algebras from the ambitwistor string worldsheet}

\author[a]{Wei Bu}

\affiliation[a]{School of Mathematics and Maxwell Institute for Mathematical Sciences\\
University of Edinburgh, EH9 3FD, UK}

\emailAdd{w.bu@sms.ed.ac.uk}

\abstract{Using the ambitwistor string, we complete the list of celestial OPE coefficients for supersymmetric theories. This uses the ambitwistor string worldsheet CFT to dynamically generate the OPE coefficients for maximally supersymmetric gauge theory, gravity and Einstein-Yang-Mills theories, including all helicity and orientation configurations. This extends previous purely bosonic results \cite{Adamo:2021zpw} to include supersymmetry and provides explicit formulas which are, to the best of our knowledge, not in the literature. We also examine how the supersymmetric infinite dimensional soft algebras behave compared to the purely bosonic cases.}

\begin{document} 
\maketitle
\flushbottom

\section{Introduction}
By Mellin transforming dynamical observables such as massless scattering amplitudes in a momentum basis, one obtains scattering amplitudes in a conformal primary basis. Since the action of the Lorentzian group on equal time slices of $\scri$, i.e. the celestial sphere, is the Mobius transformation PSL(2,$\mathbb{C}$), the Mellin transformed scattering amplitudes transform like correlation functions in conformal field theories on the celestial sphere \cite{Pasterski:2016qvg,Pasterski:2017kqt,Pasterski:2017ylz,Fan:2019emx,Banerjee:2020kaa,Banerjee:2020zlg,Banerjee:2020vnt,Guevara:2021abz,Himwich:2021dau,Jiang:2021ovh}. 

A natural question one could ask is whether this is purely coincidental or the scattering amplitudes in conformal primary basis can indeed be understood as correlators in a special CFT. For any ordinary conformal field theory, the operator spectrum and operator product expansion (OPE) are two most important characteristics. The particle spectrum of the conjectured Celestial CFT (CCFT) consists of conformal primaries. OPEs between such conformal primaries are universal features that capture the singular behavior when two operators are inserted close to each other, and the OPE coefficients are crucial to determine dynamical properties of CCFT. Some of the CCFT OPE coefficients have been computed through Mellin transform or symmetries in the literature \cite{Pate:2019lpp,Jiang:2021xzy,Fotopoulos:2020bqj}.  

Moreover, one also needs to consider how the properties of the momentum basis scattering amplitudes in the bulk are interpreted in the CCFT framework, if this correspondence were to be understood as a kind of holography. For example, in the case of soft theorems [15-17], it was observed that by reorganizing the CCFT OPEs, infinite towers of soft symmetries can be expressed in terms of infinite-dimensional algebras. For gravity and Einstein-Yang-Mills theory, the governing algebra turns out to be the extension of the Viraroso algebra, the $w_{1+\infty}$ algebra \cite{Strominger:2021lvk,Himwich:2021dau,Ahn:2021erj}. Recent literature also witnessed efforts to explore the quantum extensions of this algebra, where Einstein gravity one-loop amplitudes in the MHV sector were considered \cite{Ball:2021tmb,Mago:2021wje}. 

Besides the purely bosonic cases, supersymmetric celestial amplitudes have been explored in the literature using similar Mellin transform methods  \cite{Brandhuber:2021nez,Hu:2021lrx,Ferro:2021dub,Jiang:2021xzy}. Although celestial OPE coefficients in supersymmetric theories have been explored \cite{Jiang:2021xzy,Fotopoulos:2020bqj}, these studies have been restricted to minimal supersymmety or pure $\mathcal{N}=4$ super Yang-Mills (SYM). To our knowledge, the full list of celestial OPE coefficients for all maximally supersymmetric 4d theories has not yet been determined. Besides this, the dynamical origin of the SUSY celestial OPE coefficients is also unclear. 

In \cite{Lipstein:2015rxa,Adamo:2019ipt,Adamo:2021bej,Adamo:2021lrv,Adamo:2021zpw}, the authors made attempts to study the dynamical origin of celestial OPEs and holographic symmetries using twistor theory. In particular, by using the ambitwistor string worldsheet CFT to compute celestial OPEs, \cite{Adamo:2021zpw} has successfully generated and matched CCFT OPE coefficients with all $SL(2,\mathbb{R})$ descendants computed by Pate and collaborators \cite{Pate:2019lpp}. The results were presented in "master formulas" to include all incoming-outgoing/outgoing-outgoing orientation configurations, which contained the usual Euler-Beta functions in integral form. However, all the OPEs there only concerned gluons and gravitons. Hence the purpose of this article is to generalize this mechanism to supersymmetric cases and to calculate all possible celestial OPEs for maximally supersymmetric theories in 4d. In the end we provide a complete brochure of CCFT OPE coefficients in the form of master formulas using the worldsheet CFTs of the fully supersymmetric 4d ambitwistor strings. In order to make the discussion more self-contained, we include all subtleties in the purely bosonic cases as well as additional ones present in the supersymmetric cases. It is worth notifying that other attempts to explore the dynamical origin of CCFT have been carried out in the context of string theory \cite{Jiang:2021csc}, however the formalism there requires additional artificial manipulations during the calculations, which contrasts with the naturalness of the ambitwistor string worldsheet formalism.

The fully supersymmetric ambitwistor strings in 4d are the $\mathcal{N}=4$ SYM and $\mathcal{N}=8$ supergravity (SUGRA) ambitwistor string theories \cite{Mason:2013sva,Geyer:2014fka}. Naturally one expects that these two worldsheet CFTs should generate all the $\mathcal{N}=4$ SYM and $\mathcal{N}=8$ SUGRA CCFT OPE coefficients, however, for the $\mathcal{N}=4$ EYM theory, there are no $\mathcal{N}=4$ EYM ambitwistor worldsheet theories. Nevertheless, we are still able to compute the OPE coefficients using the appropriate vertex operators from the SYM and SUGRA ambitwistor strings. Unlike twistor string theories which are chiral in the Grassman variables, the ambitwistor string breaks the manifest $SU(\mathcal{N})$ R-symmetry into $SU(\mathcal{N}/2)\times SU(\mathcal{N}/2)$ and splits the spectrum into two halves. One half lives on twistor space and the other half on dual twistor space, which manifests the ambidextrous nature of the theory.

As shown and stated in the purely bosonic calculations \cite{Adamo:2021zpw}, the ambitwistor string exhibits properties that allow it to be a natural habitat for the interpretations of CCFT. For example, without Mellin transforming scattering amplitudes in momentum basis, the OPEs between vertex operators dynamically generate the OPE coefficients in conformal primary basis. Additionally, the worldsheet integrals of ambitwistor strings inherently localize on the boundary of the moduli space, constraining the computation precisely at the collinear region without any artificial manipulation. Besides these, the infinite tower of organizing principles for soft symmetries as well as action of soft conformal primaries on hard ones can be obtained independently, without taking soft limits on the OPE coefficients. We shall see these traits present throughout the supersymmetric calculations as well. 

Apart from the remarkable properties in the bosonic cases, the supersymmetric ambitwistor string also provides additional convenience. Due to the various helicities of particles in the spectrum, one needs to be extra cautious with the Mellin conformal scaling dimension. However, we shall see that in the worldsheet theory, the number of supersymmetry and the homogeneity of the vertex operator in twistor/dual-twistor space cohomology will resolve such subtleties automatically. As in the bosonic case, we are also able to compute the supersymmetric holographic symmetries and action of soft particles on hard ones without prior knowledge on the CCFT OPE coefficients, the $w$ algebra does not differ from the purely bosonic case, still acting as a Poisson diffeomorphism on a plane in twistor space \cite{Adamo:2021lrv}. This also matches recent results in \cite{Ahn:2021erj}. 

The paper is organized as follows, section \ref{set_up} introduces notations and basic knowledge of the ambitwistor string, section \ref{LHO} gives detailed procedure to compute all like helicity OPEs, section \ref{MHO} presents that for all mixed helicity OPEs and section \ref{SSS} concludes with the holographic symmetries and soft-hard OPEs.

\section{Set up}\label{set_up}

\subsection{Kinematics}\label{kinematics}
The study of the OPE coefficients of CCFT involves examining celestial conformal primaries inserted on the celestial sphere. It was observed that the OPE limit of CCFT coincides with the collinear limit of momentum basis scattering amplitude. In order to make this observation, we parametrize null four-momenta $k^{\mu}$ of massless particles in the following way using the stereographic coordinates $(z,\bar z)$ on the celestial sphere:
\begin{equation}
    p^{\mu} = \frac{\omega}{2}\left(1+|z|^2 ,-z-\bar z,-\im(z-\bar z),1-|z|^2\right) ,
\end{equation}
where $\omega$ denotes the energy of the particle, which when Mellin transformed becomes the conformal scaling dimension $\Delta$. Some simple algebra reveals $p_i^{\mu}(z_i,\bar z_i)\cdot p_{j,\mu}(z_j,\bar z_j) \propto |z_i-z_j|^2$, suggesting that when the two momenta become collinear as their corresponding celestial coordinates $(z_i,\bar z_i)\shortrightarrow (z_j,\bar z_j)$. The point we shall try to establish in this paper is that this clash of $(z_i,\bar z_i)\shortrightarrow (z_j,\bar z_j)$ naturally corresponds to the clash between insertions $\sigma_i\shortrightarrow \sigma_j$ of the vertex operator on the ambitwistor string worldsheet. Here we remark that the limit $(z_i, \bar z_i)\shortrightarrow (z_j,\bar z_j)$ on the celestial sphere could be separately considered as the holomorphic limit $z_{ij}=z_i-z_j\shortrightarrow 0$ and the anti-holomorphic limit $\bar z_{ij}=\bar z_i-\bar z_j\shortrightarrow 0$. Since one could only consider such chiral treatment when $z_i$ and $\bar z_i$ are independent, we complexify the celestial sphere to $S^2\times S^2$ or employ $(2,2)$ signature celestial torus \cite{Atanasov:2021oyu}. 

Besides this, we also need to capture the orientation configuration of the particle using $\veps$, namely $\veps=1$ for outgoing particles and $\veps=-1$ for incoming ones. This includes an additional parameter in our parametrization
\begin{equation}
    p^{\mu} = \frac{\veps\,\omega}{2}\left(1+z\bar z ,-z-\bar z,-\im(z-\bar z),1-z\bar z\right) ,
\end{equation}
Using spinor helicity notations, we have the following identities for our null four-momenta
\begin{align}
   p^{\alpha\Dot{\alpha}}=\sigma^{\alpha\Dot{\alpha}}_{\mu}p^{\mu} = k^{\alpha}\tilde k^{\Dot{\alpha}} &= \veps\omega \begin{pmatrix}
          1 \\
          z 
         \end{pmatrix} 
         \begin{pmatrix}
           1 & \bar z
         \end{pmatrix} ,
\end{align}
which allows us to replace spinor helicity variables $k^{\alpha}$, $\tilde k^{\Dot{\alpha}}$ with holomorphic and anti-holomorphic coordinates on the celestial sphere. More precisely, we have 
\begin{align}
    k^{\alpha} = \sqrt{\omega} \begin{pmatrix}
      1 \\
      z
    \end{pmatrix}
    \qquad
    \tilde k^{\Dot{\alpha}} = \veps\sqrt{\omega} \begin{pmatrix}
      1 \\
      \bar z
    \end{pmatrix} ,
\end{align}
Since the focus of this paper is to compute celestial OPEs for maximally supersymmetric theories, it is convenient to introduce Grassmann coordinates $\eta^A$ and their complex conjugates $\tilde\eta_A$ with designated helicities $\pm \frac{1}{2}$ on the celestial sphere, where $A$ labels R-symmetry and runs from $A=1,2,...,\mathcal{N}$. With this, we could write down supermultiplet containing all the particle content we wish to use \cite{henn2014scattering}:
\begin{multline}\label{super_expansion}
    \cU(k,\tilde k,\eta^A) = \cU_0(k,\tilde k) + \eta^{A}\cU_{1,A}(k,\tilde k) +\frac{1}{2}\eta^A\eta^B \cU_{2,AB}(k,\tilde k)+ ...\\
    + \frac{1}{(\mathcal{N}/2)!} \eta^{A_1}...\,\eta^{A_{\mathcal{N}/2}}\,\cU_{\frac{\mathcal{N}}{2},A_1...A_{\mathcal{N}/2}}(k,\tilde k)\,,
\end{multline}
where we stop at the $\cU_{\mathcal{N}/2}$ term to include half of the particle content originating from the positive helicity multiplet. The other half comes from the negative helicity multiplet:
\begin{multline}
    \bar\cU(k,\tilde k,\tilde\eta_A) = \bar\cU_0(k,\tilde k) + \tilde\eta_{A}\,\bar\cU^A_{1}(k,\tilde k) +\frac{1}{2}\tilde\eta_A\tilde\eta_B\, \bar\cU^{AB}_{2}(k,\tilde k)+ ... \\
    + \frac{1}{(\mathcal{N}/2)!} \tilde\eta_{A_1}...\,\tilde\eta_{A_{\mathcal{N}/2}}\,\bar\cU^{A_1...A_{\mathcal{N}/2}}_{\frac{\mathcal{N}}{2}}(k,\tilde k)\,,
\end{multline}
Note that this construction differs from the usual convention in the literature, where a single multiplet generates the entire spectrum. By splitting the spectrum on two multiplets, we have chosen to break the manifest $SU(\mathcal{N})$ R-symmetry and opted for $SU(\frac{\mathcal{N}}{2})\times SU(\frac{\mathcal{N}}{2})$, where $\mathcal{N}=4$ or $8$ for our purpose. Although such construction is unusual in the literature, the closest analogue can be found in  \cite{Adamo:2014yya,Adamo:2015fwa,Cachazo:2013zc}. We shall come back to this point again when we introduce all the on-shell superfield ambitwistor vertex operators in the next subsection, where $(k,\tilde k,\tilde\eta_A)$ will be assigned as the super coordinate system on the dual space. 

Notice that as all particles have different helicities and descend by $\frac{1}{2}$ as one steps down the SUSY ladder, to balance the helicities of the terms in \eqref{super_expansion}, $\eta$ shall be designated to have helicity $\frac{1}{2}$. Formally, we could define the following helicity operator:
\begin{equation}
    \mathfrak{h} := \frac{1}{2}\Big(-k^{\alpha}\partial_{k^{\alpha}}+\tilde k^{\Dot{\alpha}}\partial_{\tilde k^{\dal}}+\eta^A\partial_{\eta^A} \Big) ,
\end{equation}
If the helicity of the first bosonic particle $\cU_0$ in the multiplet has helicity $h$, the helicity operator $\mathfrak{h}$ assigns helicity $h$ to the rest of the particles in the multiplet: $\mathfrak{h}\,\cU(k,\tilde k,\eta^A)=h\,\cU(k,\tilde k,\eta^A)$, where $\cU(k,\tilde k,\eta^A)$ represents any particle in the positive helicity multiplet. Similarly, a helicity operator can be defined for particles in the negative helicity multiplet: 
\begin{equation}
    \Bar{\mathfrak{h}} := 
    \frac{1}{2}\Big(-k^{\alpha}\partial_{k^{\alpha}}+\tilde k^{\Dot{\alpha}}\partial_{\tilde k^{\dal}}+\tilde\eta_A\partial_{\tilde\eta_A} \Big) 
\end{equation}
which assigns helicity $\bar h$ to each particle originating from the negative helicity multiplet: $\Bar{\mathfrak{h}}\,\bar\cU(k,\tilde k,\tilde\eta_A)=\bar h\,\bar\cU(k,\tilde k,\tilde\eta_A)$.


\subsection{Ambitwistor string}\label{ambitwistor}
As mentioned before, we shall attempt to utilise the worldsheet CFT of the ambitwistor string to generate CCFT OPE coefficients. Hence we first introduce the tool of ambitwistor string here \cite{Mason:2013sva,Geyer:2014fka,Adamo:2013tsa}. 

Ambitwistor strings are holomorphic maps from closed Riemann surfaces to the projective ambitwistor space $\mathbb{PA}$, i.e. the supersymmetric extension of the space of complex null geodesics considered up to scale \cite{Geyer:2014lca,Lebrun:1983pa}. In four dimensions, $\mathbb{PA}$ is parametrized by twistor and dual-twistor variables ambidextrously. Together with SUSY, we have 
\begin{align}
    & \mathcal{Z}  = (\mu^{\Dot{\alpha}},\lambda_{\alpha},\chi_A)\in \mathbb{PT} \\
    & \mathcal{W}  = (\tilde\lambda_{\Dot{\alpha}},\tilde\mu^{\alpha},\tilde\chi^A) \in \mathbb{PT}^* \,,
\end{align}
where $\mathcal{Z}$ represents homogeneous coordinate on $\mathbb{CP}^{3|\mathcal{N}}$ and $\mathbb{PT}= \left\{\mathcal{Z}\in\mathbb{CP}^{3|\mathcal{N}}|\,\lambda_{\alpha}\neq 0 \right\}$. Similarly $\mathcal{W}$ denotes homogeneous coordinate on dual twistor space $\mathbb{PT}^*$. $\chi_A$ and $\tilde\chi^A$ are fermionic and $A$ ranges from $1$ to $\mathcal{N}$ labels R-symmetry. Ambitwistor space can then be represented as a quadric
\begin{equation}
    \mathbb{PA} = \left\{ (\mathcal{Z},\mathcal{W})\in \mathbb{PT}\times \mathbb{PT}^* | \mathcal{Z}\cdot \mathcal{W} =0 \right\} \,,
\end{equation}
with $\mathcal{Z}\cdot \mathcal{W} = \tilde\mu^{\alpha}\lambda_{\alpha}+\mu^{\Dot{\alpha}}\tilde\lambda_{\Dot{\alpha}}+\chi_A\tilde\chi^A$. 
Geometrically, since ambitwistor space is the complexification of all null geodesics, $\mathbb{PA}$ can be represented by the complex null geodesics and their intersection with any Cauchy surface. In the case of 4d Minkowski spacetime, $\mathbb{PA}$ is equivalent to the cotangent bundle of complexified null infinity $\mathbb{PA}\cong \mathbb{P}(T^*\scri)$ \cite{Geyer:2014lca}. Furthermore, there exist non-local relations between points in the supersymmetrized 4d Minkowski spacetime $(x,\theta,\tilde\theta)$ and a quadric $(\lambda,\tilde\lambda) \in \mathbb{CP}^1\times\mathbb{CP}^1$ in $\mathbb{PA}$:
\begin{align}
    &\mu^{\Dot{\alpha}} = \im(x^{\alpha\Dot{\alpha}}+\im \theta_A^{\alpha}\tilde\theta^{A\Dot{\alpha}})\lambda_{\alpha} \qquad \chi_A = \theta_A^{\alpha}\lambda_{\alpha} \label{incidence1} \\
    &\tilde\mu^{\alpha} = -\im(x^{\alpha\Dot{\alpha}}-\im \theta_A^{\alpha}\tilde\theta^{A\Dot{\alpha}})\tilde\lambda_{\Dot{\alpha}} \qquad \tilde\chi^A = \tilde\theta^{A\Dot{\alpha}}\tilde\lambda_{\Dot{\alpha}} \,, \label{incidence2}
\end{align}
To define a worldsheet action governing holomorphic maps from the worldsheet to $\mathbb{PA}$, one uses the worldsheet spinors $\mathcal{Z},\mathcal{W}\in \Omega^0_{\Sigma}(K_{\Sigma}^{1/2}\times\mathbb{C}^{4|\mathcal{N}})$, and a $GL(1,\mathbb{C})$ Lagrange multiplier $a \in \Omega^{0,1}_{\Sigma}$ enforcing the target space to be on the quadric $\mathcal{Z}\cdot \mathcal{W} = 0$ \cite{Geyer:2016nsh}:
\begin{equation}\label{action}
    S = \frac{1}{2\pi}\int_{\Sigma} \mathcal{W}\cdot \bar\partial \mathcal{Z}-\mathcal{Z}\cdot \bar\partial \mathcal{W}+a\mathcal{Z}\cdot \mathcal{W} +S_{\text{matter}}  \,,
\end{equation}
where $S_{\text{matter}}$ is determined by the theory one tries to describe using this action. For example, for Yang-Mills theory, $S_{\text{matter}}$ will be the action for a worldsheet current algebra $j^{\msf{a}}\in\Omega^0_{\Sigma}(K_{\Sigma}\otimes \mathfrak{g})$ for an Lie algebra $\mathfrak{g}$.
Note that the worldsheet action is invariant under any holomorphic reparametrization as well as gauge transformation associated with $GL(1,\mathbb{C})$ Lagrange multiplier $a$. Gauge fixing this redundancy using BRST procedure introduces Virasoro ghosts into the system. 

First we shall consider supersymmetric Yang-Mills theory, after gauge fixing and BRST quantization, the BRST cohomology contains vertex operators of the following form:
\begin{equation}\label{ym-vertex}
    \cU^{\msf{a}}_+(z,\bar z,\eta) = \int_{\Sigma}  \,j^{\msf{a}}(\sigma)\, a(\mathcal{Z}) , \qquad \cU^{\msf{a}}_-(z,\bar z,\tilde\eta) = \int_{\Sigma}  \,j^{\msf{a}}(\sigma)\, \tilde a(\mathcal{W}) \,,
\end{equation}
where $\sigma_i$ represents local coordinate on the worldsheet and $j^{\msf{a}}$ denotes worldsheet current of conformal weight $(1,0)$. $a(\mathcal{Z})\in H^{0,1}(\mathbb{PT},\mathcal{O})$ and $\tilde a(\mathcal{W})\in H^{0,1}(\mathbb{PT}^*,\mathcal{O})$ denote positive and negative helicity gluon wavefunctions of homogeneity degree $0$ on twistor and dual twistor space respectively. To recover spacetime free fields, we use the supersymmetrized Penrose integral formula to transform the wavefunctions $a(\mathcal{Z})$ and $\tilde a(\mathcal{W})$ \cite{Adamo:2011pv}:
\begin{align}
    &\left.\tilde F_{\Dot{\alpha}\Dot{\beta}}(x,\theta,\tilde\theta) = \int \la\lambda\d\lambda\ra\,\frac{\partial^2 a(\mathcal{Z})}{\partial \mu^{\Dot{\alpha}}\partial \mu^{\Dot{\beta}}}\right\vert_{\mu^{\Dot{\alpha}} = \im(x^{\alpha\Dot{\alpha}}+\im \theta_A^{\alpha}\tilde\theta^{A\Dot{\alpha}})\lambda_{\alpha}, \, \chi_A = \theta_A^{\alpha}\lambda_{\alpha}} \\
    &\left.F_{\alpha\beta}(x,\theta,\tilde\theta) = \int [\tilde\lambda\d\tilde\lambda]\,\frac{\partial^2 \tilde a(\mathcal{W})}{\partial \tilde\mu^{\alpha}\partial \tilde\mu^{\beta}}\right\vert_{\tilde\mu^{\alpha} = -\im(x^{\alpha\Dot{\alpha}}-\im \theta_A^{\alpha}\tilde\theta^{A\Dot{\alpha}})\tilde\lambda_{\Dot{\alpha}},\, \tilde\chi^A = \tilde\theta^{A\Dot{\alpha}}\tilde\lambda_{\Dot{\alpha}}}  \nonumber \,,
\end{align}
where the explicit form of $a(\mathcal{Z})$ and $\tilde a(\mathcal{W})$ are representation agnostic. The expressions suggest that we are restricting ourselves to the super twistor line enforced by the incidence relations in \eqref{incidence1}, \eqref{incidence2}. However, these are not the only space time field we could write down, as in \cite{Adamo:2011pv}, one could choose $\chi_A$ or $\tilde\chi^A$ to differentiate, which gives us four other space time objects:
\begin{align}
    &\left.\tilde F^A_{\Dot{\alpha}}(x,\theta,\tilde\theta) = \int \la\lambda\d\lambda\ra\,\frac{\partial^2 a(\mathcal{Z})}{\partial \mu^{\Dot{\alpha}}\partial \chi_A}\right\vert_{\mu^{\Dot{\alpha}} = \im(x^{\alpha\Dot{\alpha}}+\im \theta_A^{\alpha}\tilde\theta^{A\Dot{\alpha}})\lambda_{\alpha}, \, \chi_A = \theta_A^{\alpha}\lambda_{\alpha}} \\
    &\left.F_{\alpha A}(x,\theta,\tilde\theta) = \int [\tilde\lambda\d\tilde\lambda]\,\frac{\partial^2 \tilde a(\mathcal{W})}{\partial \tilde\mu^{\alpha}\partial \tilde\chi^A}\right\vert_{\tilde\mu^{\alpha} = -\im(x^{\alpha\Dot{\alpha}}-\im \theta_A^{\alpha}\tilde\theta^{A\Dot{\alpha}})\tilde\lambda_{\Dot{\alpha}},\, \tilde\chi^A = \tilde\theta^{A\Dot{\alpha}}\tilde\lambda_{\Dot{\alpha}}}  \nonumber \\
    &\left.\tilde F^{AB}(x,\theta,\tilde\theta) = \int \la\lambda\d\lambda\ra\,\frac{\partial^2 a(\mathcal{Z})}{\partial \chi_A\partial \chi_B}\right\vert_{\mu^{\Dot{\alpha}} = \im(x^{\alpha\Dot{\alpha}}+\im \theta_A^{\alpha}\tilde\theta^{A\Dot{\alpha}})\lambda_{\alpha}, \, \chi_A = \theta_A^{\alpha}\lambda_{\alpha}} \\
    &\left.F_{AB}(x,\theta,\tilde\theta) = \int [\tilde\lambda\d\tilde\lambda]\,\frac{\partial^2 \tilde a(\mathcal{W})}{\partial \tilde\chi^A\partial\tilde\chi^B}\right\vert_{\tilde\mu^{\alpha} = -\im(x^{\alpha\Dot{\alpha}}-\im \theta_A^{\alpha}\tilde\theta^{A\Dot{\alpha}})\tilde\lambda_{\Dot{\alpha}},\, \tilde\chi^A = \tilde\theta^{A\Dot{\alpha}}\tilde\lambda_{\Dot{\alpha}}}  \nonumber \,,
\end{align}

Together, they combine into the non-zero part of the curvature on spacetime:
\begin{align}
    &\tilde F(x,\theta,\tilde\theta) = \tilde F_{\dal\Dot{\beta}}\veps_{\alpha\beta}\,\d x^{\alpha\dal}\wedge\d x^{\beta\Dot{\beta}}+\tilde F^A_{\dal}\veps_{\alpha\beta}\,\d x^{\alpha\dal}\wedge\d\theta^{\beta}_{A}+ \tilde F^{AB}\veps_{\alpha\beta}\,\d\theta^{\alpha}_A\wedge\d\theta^{\beta}_B \\
    & F(x,\theta,\tilde\theta) =  F_{\alpha\beta}\veps_{\dal\Dot{\beta}}\,\d x^{\alpha\dal}\wedge\d x^{\beta\Dot{\beta}}+ F_{\alpha A}\veps_{\dal\Dot{\beta}}\,\d x^{\alpha\dal}\wedge\d\tilde\theta^{A\Dot{\beta}}+  F_{AB}\veps_{\dal\Dot{\beta}}\,\d\tilde\theta^{A\dal }\wedge\d\tilde\theta^{B\Dot{\beta}} \,,
\end{align}

The complete tree-level S-matrix of 4d SYM can be obtained by taking correlation functions of the vertex operators in \eqref{ym-vertex}. There are three worldsheet OPEs that need to be considered in the computations, namely the worldsheet current OPE, the $\mathcal{Z}-\mathcal{W}$ spinor OPE. First the worldsheet current OPE:
\begin{equation}
    j^{\msf{a}}(\sigma_i)j^{\msf{b}}(\sigma_j) \sim \frac{k\delta^{\msf{a}\msf{b}}}{(\sigma_i-\sigma_j)^2}\d\sigma_i\d\sigma_j+\frac{f^{\msf{a}\msf{b}\msf{c}}j^{\msf{c}}(\sigma_j)}{\sigma_i-\sigma_j}\d\sigma_i \,,
\end{equation}
where $k$ is the level of the worldsheet current algebra, $\delta^{\msf{a}\msf{b}}$ is the Killing form of the Lie group and $f^{\msf{a}\msf{b}\msf{c}}$ is the structure constant. The double pole term here comes from gravitationally mediated multi-trace interactions, here we ignore such contributions and decouple gravitational degree of freedom by setting $k \shortrightarrow 0$ \cite{Berkovits:2004jj,Adamo:2018hzd,Azevedo:2017lkz}. From now on, the worldsheet current OPE takes the following simple form:
\begin{equation}\label{jj OPE}
    j^{\msf{a}}(\sigma_i)j^{\msf{b}}(\sigma_j) \sim \frac{f^{\msf{a}\msf{b}\msf{c}}j^{\msf{c}}(\sigma_j)}{\sigma_i-\sigma_j}\d\sigma_i \,,
\end{equation}

Another OPE that needs to be accounted for is the $\mathcal{Z}-\mathcal{W}$ OPE between worldsheet spinors:
\begin{equation}\label{Z-W OPE}
    \mathcal{Z}^{I}(\sigma_i) \mathcal{W}_{J}(\sigma_j) \sim \frac{\delta^{I}_{J}\sqrt{\d\sigma_i\d\sigma_j}}{\sigma_i-\sigma_j} \,,
\end{equation}

Instead of SYM, one could also consider super gravity described by our worldsheet action \eqref{action}, which requires an additional fermionic $\rho-\tilde\rho$ system in the $S_{\text{matter}}$ term. 
\begin{align}
    & \cV_+(z,\bar z,\eta) = \int_{\Sigma}\,\left[\tilde\lambda,\frac{\partial h(\mathcal{Z})}{\partial \mu}\right]+\tilde\rho^{\Dot{\alpha}}\rho^{\Dot{\beta}} \frac{\partial^2h(\mathcal{Z})}{\partial\mu^{\Dot{\alpha}}\partial\mu^{\Dot{\beta}}} \label{integrated+} \\
    & \cV_-(z,\bar z,\tilde\eta) = \int_{\Sigma}\,\left\langle \lambda,\frac{\partial\tilde h(\mathcal{W})}{\partial \tilde\mu}\right\rangle + \rho^{\alpha}\tilde\rho^{\beta} \frac{\partial^2\tilde h(\mathcal{W})}{\partial\tilde\mu^{\alpha}\partial\tilde\mu^{\beta}} \label{integrated-}  \,,
\end{align}
where $h(\mathcal{Z})\in H^{0,1}(\mathbb{PT},\mathcal{O}(2))$ and $\tilde h(\mathcal{W})\in H^{0,1}(\mathbb{PT}^*,\mathcal{O}(2))$ are representatives of cohomology class of homogeneity degree 2 on twistor and dual twistor space respectively. Just as in the SYM case, we use the supersymmetric version of the Penrose transform to obtain momentum eigenstates on spacetime \cite{Adamo:2011pv}. 
The correlation functions of these vertex operators correctly produce the entire tree-level S-matrix of Einstein super gravity, besides the $\mathcal{Z}-\mathcal{W}$ OPE in \eqref{Z-W OPE} we just introduced, an additional $\rho-\tilde\rho$ OPE needs to be accounted for in such computations:
\begin{equation}\label{rho-rho_OPE}
    \rho^{I}(\sigma_i)\tilde\rho_J(\sigma_j) \sim \frac{\delta^{I}_{J}\sqrt{\d\sigma_i\d\sigma_j}}{\sigma_i-\sigma_j}
\end{equation}

\subsection{Vertex operators in conformal primary basis}\label{vertex_operator}
Now as we mentioned before, the vertex operators of our worldsheet theory can be written in any basis, however, to make contact with the existing literature in the celestial holography community, we shall adopt the conformal primary basis here. 

We shall start with the gluon wavefunctions here \cite{Adamo:2019ipt}:
\begin{align}
    & a(\mathcal{Z}) = \int_{\mathbb{C}^*\times\mathbb{R}_+}\,\frac{\d s}{s}\,\frac{\d t}{t^{2-\Delta}}\,\bar\delta^2(z-s\lambda(\sigma))\,e^{\im s\,t\,\veps[\mu(\sigma)\bar z]+\im s\,\sqrt{t}\,\chi_A(\sigma)\tilde\eta^{A}}  \label{sym_multiplet_+}\\ 
    & \tilde a(\mathcal{W}) = \int_{\mathbb{C}^*\times \mathbb{R}_+}\,\frac{\d s}{s}\,\frac{\d t}{t^{2-\Delta}}\bar\delta^2(\veps \bar z-s\tilde\lambda(\sigma))\,e^{\im s\,t\,\langle\tilde\mu(\sigma)z\rangle+\im s\,\sqrt{t}\,\tilde\chi^A(\sigma)\eta_{A}}  \label{sym_multiplet_-} \,,
\end{align}
where $z_{\alpha}=(1,z)$, $\bar z_{\Dot{\alpha}}=(1,\bar z)$ as we have in the kinematics section. The holomorphic delta functions or the scattering equations are defined as follows \cite{Cachazo:2013hca,Cachazo:2014xea}
\begin{equation}
    \bar\delta^2(z-s\lambda(\sigma)) := \frac{1}{(2\pi\im)^2} \bigwedge_{\alpha = 0,1} \bar\partial\left(\frac{1}{z_{\alpha}-s\lambda_{\alpha}(\sigma)} \right) \,,
\end{equation}
One could check that it indeed acts as a delta function enforcing the content inside its brackets to vanish. Notice that the position of $\veps$ in the negative helicity wavefunction is slightly different compared to the one used in \cite{Adamo:2021zpw}. Instead of placing it on the exponential, it sits in front of $\bar z$. Rigorously speaking, this would contribute an overall sign factor in front when fermions are involved in the calculation. Since \cite{Adamo:2021zpw} only considered bosonic particles, this was not an issue there. However, for the supersymmetric theories we consider, it is important to place the orientation parameter $\veps$ in front of $\bar z$ at all times.

As the expansion suggested by equation \eqref{super_expansion}, to extract individual vertex operators from the gluon wavefunctions \eqref{sym_multiplet_+} and \eqref{sym_multiplet_-}, we take derivatives with respect to $\eta$ or $\tilde\eta$ from the factors $e^{\im s\sqrt{t}\,\tilde\chi^A(\sigma)\eta_{A}}$ or $e^{\im s\sqrt{t}\,\chi_A(\sigma)\tilde\eta^A}$ in the vertex operators to obtain the desired the number of $\eta$ or $\tilde\eta$. 
\begin{align}
    &\cU^{\msf{a}}_{+,\Delta}(z,\bar z,\eta^A) = \cO^{\msf{a}}_{+,\Delta}(z,\bar z, h)+\eta^A\Gamma^{\msf{a}}_{+,\Delta,A}(z,\bar z)+\frac{1}{2}\eta^A\eta^B\Phi^{\msf{a}}_{\Delta,AB}(z,\bar z) \\
    &\cU^{\msf{a}}_{-,\Delta}(z,\bar z,\tilde\eta_A) = \cO^{\msf{a}}_{-,\Delta}(z,\bar z, h)+\tilde\eta_A\bar\Gamma^{\msf{a},A}_{-,\Delta}(z,\bar z)+\frac{1}{2}\tilde\eta_A\tilde\eta_B\Phi^{\msf{a},AB}_{\Delta}(z,\bar z) \,,
\end{align}
We indeed see that all particles have the same conformal weight since the Grassman variables have weight $0$. Due to the designated helicities of $\eta$ and $\tilde\eta$ being $\pm \frac{1}{2}$, our particles also have the right helicity. However, we notice that, unlike usual SUSY expansions, negative helicity particles cannot be generated from the positive helicity multiplet nor vice versa. This suggests that the $SU(4)$ R-symmetry is not manifest in ambitwistor space $\mathbb{PA}$, it splits between twistor and dual twistor space, reflecting the ambidextrous nature of our theory.
Here we summerize the particle content in $\mathcal{N}=4$ SYM.

\begin{table}[H]
    \centering
    \begin{tabular}{c|c|c|c}
       Particle  & $\cO^{\msf{a}}$ & $\Gamma^{\msf{a}}_A$ & $\Phi^{\msf{a}}_{AB}$ \\
       \hline
       Helicity    &  $+1$  & $+\frac{1}{2}$ & $0$
    \end{tabular}
\end{table}
\noindent from the positive helicity supermultiplet \eqref{sym_multiplet_+}, and 

\begin{table}[H]
    \centering
    \begin{tabular}{c|c|c|c}
       Particle  & $\cO^{\msf{a}}$ & $\bar\Gamma^{\msf{a},A}$ & $\Phi^{\msf{a},AB}$ \\
       \hline
       Helicity    &  $-1$  & $-\frac{1}{2}$ & $0$
    \end{tabular}
\end{table}
\noindent from the negative helicity supermultiplet \eqref{sym_multiplet_-}.
The first particles we extract are the spin $1$ gluons of both $\pm$ helicities:
\begin{align}
    \cO^{\msf{a},\veps}_{+,\Delta}(z,\bar z) =\int_{\Sigma}\, j^{\msf{a}}(\sigma) \int_{\mathbb{C}^*\times\mathbb{R}_+}\,\frac{\d s}{s}\,\frac{\d t}{t^{2-\Delta}}\,&\bar\delta^2(z-s\lambda(\sigma))\,\exp\Big(\im\veps\,t\,s\,[\mu(\sigma)\,\bar z]\Big)  \\ 
    \cO^{\msf{a},\veps}_{-,\Delta}(z,\bar z) = \int_{\Sigma}\, j^{\msf{a}}(\sigma)\int_{\mathbb{C}^*\times \mathbb{R}_+}\,\frac{\d s}{s}\,\frac{\d t}{t^{2-\Delta}}&\bar\delta^2(\veps\,\bar z-s\tilde\lambda(\sigma))\,\exp\Big(\im t\,s\,\la\tilde\mu(\sigma)\, z\ra\Big) \,,
\end{align}
where we read off the term with zeroth power in $\eta$ from both positive and negative helicity multiplets and set $\eta=\tilde\eta=0$. Note that this is exactly the same vertex operator we used in \cite{Adamo:2021zpw} to compute bosonic OPEs. Similarly, if we extract the operators with first order in $\eta$ or $\tilde\eta$ and set $\eta=\tilde\eta=0$, we have the spin $\frac{1}{2}$ gluinos:
\begin{align}
    \Gamma^{\msf{a},\veps}_{+,\Delta,A}(z,\bar z) =\int_{\Sigma}\, j^{\msf{a}}(\sigma) \int_{\mathbb{C}^*\times\mathbb{R}_+}\,\frac{\d s}{s}\,\frac{\d t}{t^{\frac{3}{2}-\Delta}}\,&\bar\delta^2(z-s\lambda(\sigma))\,\left(\im s\chi_A(\sigma) \right)
    \,\exp\Big(\im\veps\,t\,s\,[\mu(\sigma)\,\bar z]\Big)  \\ 
    \bar\Gamma^{\msf{a},\veps,A}_{-,\Delta}(z,\bar z) = \int_{\Sigma}\, j^{\msf{a}}(\sigma)\int_{\mathbb{C}^*\times \mathbb{R}_+}\,\frac{\d s}{s}\,\frac{\d t}{t^{\frac{3}{2}-\Delta}}&\bar\delta^2(\veps\,\bar z-s\tilde\lambda(\sigma))\,\left(\im s\tilde\chi^A(\sigma) \right)\,\exp\Big(\im\,t\,s\,\la\tilde\mu(\sigma)\, z\ra\Big)  \,,
\end{align}
Notice that the power of $t$ differs from the gluon, this is reflecting the helicity of the particle we are writing down, the Mellin conformal scaling dimension varies as the helicity of the particle varies. This is taken care of by the $\sqrt{t}$ factor on the exponential. 

It is worth noting that the homogeneities of the corresponding vertex operators are not the same, since factors of $s$ were brought down through differentiation. For example, the positive helicity gluinos here are of homogeneity $-1$ on twistor space and the negative helicity ones are of homogeneity $-1$ on dual twistor space.

The last particle in the $\mathcal{N}=4$ SYM spectrum is the scalar, which just amounts to taking the term with second order in $\eta$ or $\tilde\eta$ and setting $\eta=\tilde\eta=0$.
\begin{align}
    \Phi^{\msf{a},\veps}_{\Delta,AB}(z,\bar z) =\int_{\Sigma}\, j^{\msf{a}}(\sigma) \int\,\frac{\d s}{s}\,\frac{\d t}{t^{1-\Delta}}\,&\bar\delta^2(z-s\lambda(\sigma))\, 
    \left(-s^2\chi_A(\sigma)\chi_B(\sigma) \right)\,
    \exp\Big(\im\veps\,t\,s\,[\mu(\sigma)\,\bar z]\Big) \label{scalar+} \\ 
    \Phi^{\msf{a},\veps,AB}_{\Delta}(z,\bar z) = \int_{\Sigma}\, j^{\msf{a}}(\sigma)\int\,\frac{\d s}{s}\,\frac{\d t}{t^{1-\Delta}}&\bar\delta^2(\veps\,\bar z-s\tilde\lambda(\sigma))\,\left(- s^2\tilde\chi^A(\sigma)\tilde\chi^B(\sigma) \right)\,\exp\Big(\im\,t\,s\,\la\tilde\mu(\sigma)\, z\ra\Big) \,, \label{scalar-}
\end{align}
Notice that there are two different representations of the scalar originating from either the positive or negative helicity gluon wavefunction. 
This precisely reflects the fact that we are in ambitwistor space, where we have chosen to break the manifest $SU(4)$ symmetry of $\mathcal{N}=4$ SYM into $SU(2)\times SU(2)$. Hence the two representations each takes half of the scalars. \eqref{scalar+} and \eqref{scalar-} are related by $SU(4)$ transformation, which is not manifest in ambitwistor space.
Later in the computations, we shall observe that in order to obtain vertex operators with correct homogeneity, one is forced to stay ambidextrous and use both of these representations of the scalar when needed. 

Next we write down vertex operators for particles in the $\mathcal{N}=8$ supergravity multiplet. Firstly the wavefunctions $h(\mathcal{Z})$ and $\tilde h(\mathcal{W})$ in \eqref{integrated+} and \eqref{integrated-} in conformal primary basis can be written as 
\begin{align}
    & h(\mathcal{Z}) = \int_{\mathbb{C}^*\times\mathbb{R}_+} \frac{\d s}{s^3}\frac{\d t}{t^{3-\Delta}}\bar\delta^2(z-s\lambda)\,e^{\im\,\veps\, t\, s\,[\mu(\sigma)\,\bar z]+\im s\,\sqrt{t}\,\chi_A(\sigma)\tilde\eta^{A}}\label{sugra_multiplet_+}  \\ 
    & \tilde h(\mathcal{W})= \int_{\mathbb{C}^*\times\mathbb{R}_+}\,\frac{\d s}{s^3}\,\frac{\d t}{t^{3-\Delta}}\,\bar\delta^2(\veps\,\bar z-s\tilde\lambda(\sigma))\,e^{\im\, s\,t\,\langle\tilde\mu(\sigma)\,z\rangle+\im s\,\sqrt{t}\,\tilde\chi^A(\sigma)\eta_{A}} \label{sugra_multiplet_-} \,,
\end{align}
The power of $s$ differs from the gluon multiplet as the cohomological homogeneity degree of the graviton wavefunction is $2$ instead of $0$. Notice that the position of $\veps$ in the negative helicity wavefunction also differs from that in \cite{Adamo:2021zpw} for reasons we explained in the gluon case. Following similar expansions as the SYM case, we have from the positive helicity graviton supermultiplet \eqref{sugra_multiplet_+}:
\begin{table}[H]
    \centering
    \begin{tabular}{c|c|c|c|c|c}
       Particle  & $\cG$ & $\Theta_A$ & $V_{AB}$ & $\Xi_{ABC}$ & $\Pi_{ABCD}$ \\
       \hline
       Helicity  &  $+2$ & $+\frac{3}{2}$ & $+1$  & $+\frac{1}{2}$ & $0$ 
    \end{tabular}
\end{table}
\noindent and from the negative helicity graviton supermultiplet \eqref{sugra_multiplet_-}:
\begin{table}[H]
    \centering
    \begin{tabular}{c|c|c|c|c|c}
       Particle  & $\cG$ & $\bar\Theta^A$ & $\bar V^{AB}$ & $\bar\Xi^{ABC}$ & $\Pi^{ABCD}$ \\
       \hline
       Helicity  &  $-2$ & $-\frac{3}{2}$ & $-1$  & $-\frac{1}{2}$ & $0$ 
    \end{tabular}
\end{table}
\noindent After substituting $h(\mathcal{Z})$ and $\tilde h(\mathcal{W})$ in \eqref{integrated+} and \eqref{integrated-}, we extract the terms with zeroth power in $\eta$ or $\tilde\eta$ and set $\eta=\tilde\eta=0$, the spin $2$ gravitons can be presented as follows:
\begin{align}
\cG^{\veps}_{+,\Delta}(z,\bar z)=  \veps\int\frac{\d s}{s^2}\,\frac{\d t}{t^{2-\Delta}}\, &
\,\Big(\im[\tilde\lambda(\sigma)\bar z]-\veps\, s\,t[\rho(\sigma)\bar z][\tilde\rho(\sigma)\bar z] \Big) \nonumber\\
&\bar\delta^2\left(z-s\,\lambda(\sigma)\right) \,\exp\Big(\im\veps\,t\,s\,[\mu(\sigma)\,\bar z]\Big) \\
\cG^{\veps}_{-,\Delta}(z,\bar z)= \int\frac{\d s}{s^2}\,\frac{\d t}{t^{2-\Delta}}\, &
\,\Big(\im\la\lambda(\sigma) z\ra-s\,t\,\la\tilde\rho(\sigma) z\ra\la\rho(\sigma)z\ra \Big) \nonumber\\
&\bar\delta^2\left(\veps\,\bar z-s\,\tilde\lambda(\sigma)\right) \,\exp\Big(\im\,t\,s\,\la\tilde\mu(\sigma)\, z\ra\Big) \,,
\end{align}
To go one step down the supermultiplets, we extract the vertex operator with one power of $\eta$ or $\tilde\eta$ describing the spin $\pm\frac{3}{2}$ gravitinos, after setting $\eta=\tilde\eta =0$:
\begin{align}
\Theta^{\veps}_{+,\Delta,A}(z,\bar z)=  \veps\int\frac{\d s}{s^2}\,\frac{\d t}{t^{\frac{3}{2}-\Delta}}\,&
\,\Big(\im[\tilde\lambda(\sigma)\bar z]-\veps\,s\,t\,[\rho(\sigma)\bar z][\tilde\rho(\sigma)\bar z] \Big)\left(\im s\chi_A(\sigma)\right) \nonumber\\
&\,\bar\delta^2\left(z-s\,\lambda(\sigma)\right) \,\exp\Big(\im\veps\,t\,s\,[\mu(\sigma)\,\bar z]\Big) \\
\bar\Theta^{\veps,A}_{-,\Delta}(z,\bar z)=  \int\frac{\d s}{s^2}\,\frac{\d t}{t^{\frac{3}{2}-\Delta}}\,&
\,\Big(\im\la\lambda(\sigma) z\ra-s\,t\,\la\tilde\rho(\sigma) z\ra\la\rho(\sigma)z\ra \Big)\left(\im s\tilde\chi^A(\sigma)\right)\, \nonumber\\
&\bar\delta^2\left(\veps\bar z-s\,\tilde\lambda(\sigma)\right) \,\exp\Big(\im\,t\,s\,\la\tilde\mu(\sigma)\, z\ra\Big) \,,
\end{align}
Similarly we could obtain the spin $1$ gauge bosons:
\begin{align}
V^{\veps}_{+,\Delta,AB}(z,\bar z)=  \veps\int\frac{\d s}{s^2}\,&\frac{\d t}{t^{1-\Delta}}\, \,\Big(\im[\tilde\lambda(\sigma)\bar z]-\veps\,s\,t\,[\rho(\sigma)\bar z][\tilde\rho(\sigma)\bar z] \Big)\,\left(- s^2\chi_A(\sigma)\chi_B(\sigma)\right) \nonumber\\
&\,\bar\delta^2\left(z-s\,\lambda(\sigma)\right) \,\exp\Big(\im\veps\,t\,s\,[\mu(\sigma)\,\bar z]\Big) \\
\bar V^{\veps,AB}_{-,\Delta}(z,\bar z)=  \int\frac{\d s}{s^2}\,&\frac{\d t}{t^{1-\Delta}}\,
\,\Big(\im\la\lambda(\sigma) z\ra-s\,t\,\la\tilde\rho(\sigma) z\ra\la\rho(\sigma)z\ra \Big)\,\left(- s^2\tilde\chi^A(\sigma)\tilde\chi^B(\sigma)\right) \nonumber\\
&\,\bar\delta^2\left(\veps\,\bar z-s\,\tilde\lambda(\sigma)\right) \,\exp\Big(\im\,t\,s\,\la\tilde\mu(\sigma)\, z\ra\Big) \,,
\end{align}
and the spin $\frac{1}{2}$ gauginos:
\begin{align}
\Xi^{\veps}_{+,\Delta,ABC}(z,\bar z)=  \veps\int\frac{\d s}{s^2}\,\frac{\d t}{t^{\frac{1}{2}-\Delta}}\,&
\,\Big(\im[\tilde\lambda(\sigma)\bar z]-\veps\,s\,t[\rho(\sigma)\bar z][\tilde\rho(\sigma)\bar z] \Big)\,\nonumber \\
\left(-\im s^3\chi_A(\sigma)\chi_B(\sigma)\chi_C(\sigma)\right) &\,\bar\delta^2\left(z-s\,\lambda(\sigma)\right) \,\exp\Big(\im\veps\,t\,s\,[\mu(\sigma)\,\bar z]\Big) \\
\Xi^{\veps,ABC}_{-,\Delta}(z,\bar z)=  \int\frac{\d s}{s^2}\,\frac{\d t}{t^{\frac{1}{2}-\Delta}}\,&
\,\Big(\im\la\lambda(\sigma) z\ra-s\,t\,\la\tilde\rho(\sigma) z\ra\la\rho(\sigma)z\ra \Big)\, \nonumber\\
\left(-\im s^3\tilde\chi^A(\sigma)\tilde\chi^B(\sigma)\tilde\chi^C(\sigma)\right)\,&\bar\delta^2\left(\veps\,\bar z-s\,\tilde\lambda(\sigma)\right) \,\exp\Big(\im\,t\,s\,\la\tilde\mu(\sigma)\, z\ra\Big) \,,
\end{align}
as well as the scalars:
\begin{align}
\Pi^{\veps}_{\Delta,ABCD}(z,\bar z)&=  \veps\int\frac{\d s}{s^2}\,\frac{\d t}{t^{-\Delta}}\,
\,\Big(\im[\tilde\lambda(\sigma)\bar z]-\veps\,s\,t\,[\rho(\sigma)\bar z][\tilde\rho(\sigma)\bar z] \Big)\,\nonumber\\
&\left(s^4\chi_A(\sigma) \chi_B(\sigma)\chi_C(\sigma)\chi_D(\sigma)\right)\,\bar\delta^2\left(z-s\,\lambda(\sigma)\right) \,\exp\Big(\im\veps\,t\,s\,[\mu(\sigma)\,\bar z]\Big) \label{scalar1+} \\
\Pi^{\veps,ABCD}_{\Delta}(z,\bar z)&=  \int\frac{\d s}{s^2}\,\frac{\d t}{t^{-\Delta}}\,
\,\Big(\im\la\lambda(\sigma) z\ra-s\,t\,\la\tilde\rho(\sigma) z\ra\la\rho(\sigma)z\ra \Big)\,\nonumber\\
&\left(s^4\tilde\chi^A(\sigma)\tilde\chi^B(\sigma)\tilde\chi^C(\sigma)\tilde\chi^D(\sigma)\right)\,\bar\delta^2\left(\veps\,\bar z-s\,\tilde\lambda(\sigma)\right) \,\exp\Big(\im\,t\,s\,\la\tilde\mu(\sigma)\, z\ra\Big)  \label{scalar1-}
\end{align}
Note that just as in SYM, we obtain two representations of the gravity scalar from two origins. For the same reason as in SYM, the $SU(8)$ symmetry is manifestly broken into $SU(4)\times SU(4)$ in ambitwistor space, hence \eqref{scalar1+} and \eqref{scalar1-} each represents half of the gravity scalars. 

In the following two sections, we demonstrate the methodology of computing OPEs involving the gluinos, the gravitinos and their bosonic super partners in different helicity and orientation configurations. Since all other OPEs follow similar procedures, we simply list the results in the appendix.

\section{Like helicity OPEs}\label{LHO}
\subsection{Gluino-gluino OPE}\label{same_gluino_gluino}
Here we spell out the calculation explicitly for the like helicity gluino-gluino OPE. Majority of the steps here will follow through from the like helicity gluon-gluon computation in \cite{Adamo:2021zpw}, apart from the subtlety of the additional $\chi-\tilde\chi$ fermionic OPE.
\begin{multline}
\Gamma^{\msf{a},\veps_i}_{+,\Delta_i,A}(z_i,\bar z_i)\,\Gamma^{\msf{b},\veps_j}_{+,\Delta_j,B}(z_j,\bar z_j) \sim \int_{\Sigma_i\times\Sigma_j\times(\mathbb{C}^*)^2\times\mathbb{R}_+^2}\d\sigma_i\,\frac{f^{\msf{abc}}\,j^{\msf{c}}(\sigma_j)}{\sigma_{ij}}\,\frac{\d s_i}{s_i}\,\frac{\d s_j}{s_j}\,\frac{\d t_i}{t_i^{\frac{3}{2}-\Delta_i}}\,\frac{\d t_j}{t_j^{\frac{3}{2}-\Delta_j}}\,\\(\im s_i\chi_A(\sigma_i))(\im s_j\chi_B(\sigma_j))\, 
\bar\delta^2\left(z_i-s_i\,\lambda(\sigma_i)\right)\bar\delta^2\left(z_j-s_j\,\lambda(\sigma_j)\right)\,\\
\exp\Big(\im\veps_i\,t_i\,s_i\,[\mu(\sigma_i)\,\bar z_i]+\im\veps_j\,t_j\,s_j\,[\mu(\sigma_j)\,\bar z_j]\Big)\,, 
\end{multline}
where we have already performed the $j-j$ OPE given by \eqref{jj OPE}. The immediate feature of this OPE one could observe is that in the limit $\sigma_i-\sigma_j= \sigma_{ij}\shortrightarrow 0$, the two delta functions simultaneously enforce $\la z_iz_j\ra = z_i-z_j\shortrightarrow 0$. This simply remarks the fact that the collision of two vertex operator insertions on the ambitwistor string worldsheet coincides with the collision of insertion points on the celestial sphere. With this understanding in place, we could begin manipulating the expression to see the desired OPE appearing. In the following we simply do some of the integrals here to make the resulting vertex operator more manifest. 

The first thing we need to do here is to perform the $s_i$ integral against the first delta function. This sets $s_{i}=\la\xi\,z_i\ra/\la\xi\,\lambda(\sigma_i)\ra$ for arbitrary reference spinor $\xi_{\alpha} \neq z_{i\,\alpha}$. For simplicity we set $\xi_{\alpha}=\iota_{\alpha}=(0,1)$, for which $\la\iota\,z_i\ra=1$. This gives us
\begin{multline}
\Gamma^{\msf{a},\veps_i}_{+,\Delta_i,A}(z_i,\bar z_i)\,\Gamma^{\msf{b},\veps_j}_{+,\Delta_j,B}(z_j,\bar z_j) \sim \int\d\sigma_i\,\frac{f^{\msf{abc}}\,j^{\msf{c}}(\sigma_j)}{\sigma_{ij}}\,{\d s_j}\,\frac{\d t_i}{t_i^{\frac{3}{2}-\Delta_i}}\,\frac{\d t_j}{t_j^{\frac{3}{2}-\Delta_j}}\,\bar{\delta}\!\left(\la z_{i}\,\lambda(\sigma_i)\ra\right)\\
\,(-\chi_A(\sigma_i)\chi_B(\sigma_j))\,\bar\delta^2\left(z_j-s_j\,\lambda(\sigma_j)\right)\,\exp\left(\im\veps_i\,t_i\,\frac{[\mu(\sigma_i)\,\bar z_i]}{\la\iota\,\lambda(\sigma_i)\ra}+\im\veps_j\,t_j\,s_j\,[\mu(\sigma_j)\,\bar z_j]\right)\,,
\end{multline}
Notice that the number of $\chi$ remaining and the color current $j^{\msf{a}}(\sigma_j)$ already indicate that what we end up with on the right hand side should be the scalar in the gluon multiplet $\Phi^{\msf{a}}_{AB}$, the following computation will reveal whether the homogeneity of the resulting expression matches that of a scalar. Now we could use the definition of the holomorphic delta function to integrate by parts to trade $\bar\delta(\la z_i\lambda(\sigma_i)\ra)$ with the $\sigma_{ij}$ pole, obtaining
\begin{multline}
\Gamma^{\msf{a},\veps_i}_{+,\Delta_i,A}(z_i,\bar z_i)\,\Gamma^{\msf{b},\veps_j}_{+,\Delta_j,B}(z_j,\bar z_j) \sim f^{\msf{abc}}\int\d\sigma_i\,\frac{j^{\msf{c}}(\sigma_j)}{\la z_i\lambda(\sigma_i)\ra}\,\bar\delta(\sigma_{ij})\,{\d s_j}\,\frac{\d t_i}{t_i^{\frac{3}{2}-\Delta_i}}\,\frac{\d t_j}{t_j^{\frac{3}{2}-\Delta_j}}\\
(-\chi_A(\sigma_i)\chi_B(\sigma_j))\,\bar\delta^2\left(z_j-s_j\,\lambda(\sigma_j)\right)\,\exp\left(\im\veps_i\,t_i\,\frac{[\mu(\sigma_i)\,\bar z_i]}{\la\iota\,\lambda(\sigma_i)\ra}+\im\veps_j\,t_j\,s_j\,[\mu(\sigma_j)\,\bar z_j]\right)\,,
\end{multline}
Then we perform the $\sigma_i$ integral with respect to the delta function $\bar\delta(\sigma_{ij})$ to get 
\begin{multline}
\Gamma^{\msf{a},\veps_i}_{+,\Delta_i,A}(z_i,\bar z_i)\,\Gamma^{\msf{b},\veps_j}_{+,\Delta_j,B}(z_j,\bar z_j) \sim f^{\msf{abc}}\int_{\Sigma_j\times \mathbb{C}^*\times\mathbb{R}^2_+}\,\frac{j^{\msf{c}}(\sigma_j)}{\la z_i\lambda(\sigma_j)\ra}\,{\d s_j}\,\frac{\d t_i}{t_i^{\frac{3}{2}-\Delta_i}}\,\frac{\d t_j}{t_j^{\frac{3}{2}-\Delta_j}}\, \\
(-\chi_A(\sigma_j)\chi_B(\sigma_j))\,\bar\delta^2\left(z_j-s_j\,\lambda(\sigma_j)\right)\,\exp\left(\im\veps_i\,t_i\,\frac{[\mu(\sigma_j)\,\bar z_i]}{\la\iota\,\lambda(\sigma_j)\ra}+\im\veps_j\,t_j\,s_j\,[\mu(\sigma_j)\,\bar z_j]\right)\,,
\end{multline}
The remaining delta function  enforces the following relations:
\begin{equation}
    \frac{\la\iota\,\lambda(\sigma_j)\ra}{\la z_{i}\,\lambda(\sigma_j)\ra}=\frac{\la\iota\,z_j\ra}{\la z_{i}\,z_j\ra}=\frac{1}{z_{ij}}\,, \qquad \frac{1}{\la\iota\,\lambda(\sigma_j)\ra}=\frac{s_j}{\la\iota\,z_{j}\ra}=s_j\,.
\end{equation}
to obtain our pole $\frac{1}{\la z_i\lambda(\sigma_j)\ra} = \frac{s_j}{ z_{ij}}$. Substitute these in our expression, which reads
\begin{multline}
\Gamma^{\msf{a},\veps_i}_{+,\Delta_i,A}(z_i,\bar z_i)\,\Gamma^{\msf{b},\veps_j}_{+,\Delta_j,B}(z_j,\bar z_j) \sim \frac{f^{\msf{abc}}}{z_{ij}}\int\,j^{\msf{c}}(\sigma_j)\,{\d s_j}\,s_j\,\frac{\d t_i}{t_i^{\frac{3}{2}-\Delta_i}}\,\frac{\d t_j}{t_j^{\frac{3}{2}-\Delta_j}}\,(-\chi_A(\sigma_j)\chi_B(\sigma_j)) \\
\bar\delta^2\left(z_j-s_j\,\lambda(\sigma_j)\right)\,\exp\left(\im\veps_i\,t_i\,s_j\,[\mu(\sigma_j)\,\bar z_i]+\im\veps_j\,t_j\,s_j\,[\mu(\sigma_j)\,\bar z_j]\right)\,,
\end{multline}
Note that the scalar $\Phi^{\msf{a}}_{AB}$ in integral expression requires $\d s_j s_j$ to have homogeneity $-2$, which in our expression is given by the identities we just used. 

To proceed from here, one needs to combine the exponential terms. To do this, we notice that the first term in the exponential requires an additional $t_j$ factor to match the content of the second term, hence we rescale $t_i\mapsto t_it_j$ to get 
\begin{multline}
\Gamma^{\msf{a},\veps_i}_{+,\Delta_i,A}(z_i,\bar z_i)\,\Gamma^{\msf{b},\veps_j}_{+,\Delta_j,B}(z_j,\bar z_j) \sim \frac{f^{\msf{abc}}}{z_{ij}}\int\,j^{\msf{c}}(\sigma_j)\,{\d s_j}\,s_j\,\frac{\d t_i}{t_i^{\frac{3}{2}-\Delta_i}}\,\frac{\d t_j}{t_j^{2-\Delta_i-\Delta_j}}\,(-\chi_A(\sigma_j)\chi_B(\sigma_j)) \\
\,\bar\delta^2\left(z_j-s_j\,\lambda(\sigma_j)\right)\,\exp\left(\im\veps_i\,t_i\,t_j\,s_j\,[\mu(\sigma_j)\,\bar z_i]+\im\veps_j\,t_j\,s_j\,[\mu(\sigma_j)\,\bar z_j]\right)\,,
\end{multline}
From here, some algebra on the exponential leads us to
\begin{multline}
\im\,t_j\,s_j\left(\veps_i\,t_i\,[\mu(\sigma_j)\,\bar z_i]+\veps_j\,[\mu(\sigma_j)\,\bar z_j]\right) \\ =\im\,s_{j}\,t_j\left(1+\frac{\veps_i}{\veps_j}\,t_i\right)\left(\frac{\veps_i\,t_i}{1+\frac{\veps_i}{\veps_j}\,t_i}[\mu(\sigma_j)\,\bar{z}_{ij}]+\veps_j\,[\mu(\sigma_j)\,\bar{z}_j]\right)\,,
\end{multline}
where $\bar z_{ij\,\dal} := \bar z_{i\,\dal} - \bar z_{j\,\dal}$. Now we need to get rid of the dependence of $t_i$ on the exponential, which could be achieved by rescaling $t_j \mapsto t_j\,|1+\frac{\veps_i}{\veps_j}t_i|$.
\begin{multline}
\Gamma^{\msf{a},\veps_i}_{+,\Delta_i,A}(z_i,\bar z_i)\,\Gamma^{\msf{b},\veps_j}_{+,\Delta_j,B}(z_j,\bar z_j) \sim \frac{f^{\msf{abc}}}{z_{ij}}\int\,j^{\msf{c}}(\sigma_j)\,{\d s_j}\,s_j\,\frac{\d t_i\,t_i^{\Delta_i-\frac{3}{2}}}{|1+\frac{\veps_i}{\veps_j}t_i|^{\Delta_i+\Delta_j-1}}\,\frac{\d t_j}{t_j^{2-\Delta_i-\Delta_j}}\, \\
(-\chi_A(\sigma_j)\chi_B(\sigma_j))\,\bar\delta^2\left(z_j-s_j\,\lambda(\sigma_j)\right)\,\exp\left[\im\,t_j\,s_j\left(\frac{\veps_i\,t_i}{|1+\frac{\veps_i}{\veps_j}\,t_i|}\,[\mu(\sigma_j)\,\bar{z}_{ij}]+\mathrm{sgn}(\veps_j+\veps_i\,t_i)\,[\mu(\sigma_j)\,\bar z_j]\right)\right]\,,
\end{multline}
where $\mathrm{sgn}$ denotes the sign function. We notice that all $t_i$ dependence on the exponential is now bundled together with a factor of $\bar z_{ij}$, which vanishes in the OPE limit. To make the form of the integral more transparent, we rearrange to get 
\begin{multline}
\Gamma^{\msf{a},\veps_i}_{+,\Delta_i,A}(z_i,\bar z_i)\,\Gamma^{\msf{b},\veps_j}_{+,\Delta_j,B}(z_j,\bar z_j) \sim \frac{f^{\msf{abc}}}{z_{ij}}  \int_{\mathbb{R}_+} \frac{\d t_i\,t_i^{\Delta_i-\frac{3}{2}}}{|1+\frac{\veps_i}{\veps_j}t_i|^{\Delta_i+\Delta_j-1}} \int\,j^{\msf{c}}(\sigma_j)\, \\
\int_{\mathbb{C}^*\times\mathbb{R}_+}\frac{\d s_j}{s_j}\,\frac{\d t_j}{t_j^{1-(\Delta_i+\Delta_j-1)}}\,(\im s_j\chi_A(\sigma_j))(\im s_j\chi_B(\sigma_j))\bar\delta^2\left(z_j-s_j\,\lambda(\sigma_j)\right)\,\\
\exp\left[\im\,t_j\,s_j\left(\frac{\veps_i\,t_i}{|1+\frac{\veps_i}{\veps_j}\,t_i|}\,[\mu(\sigma_j)\,\bar{z}_{ij}]+\mathrm{sgn}(\veps_j+\veps_i\,t_i)\,[\mu(\sigma_j)\,\bar z_j]\right)\right]
\,,
\end{multline}
Now we just Taylor expand in the first term on the exponential to get 
\begin{multline}
\Gamma^{\msf{a},\veps_i}_{+,\Delta_i,A}(z_i,\bar z_i)\,\Gamma^{\msf{b},\veps_j}_{+,\Delta_j,B}(z_j,\bar z_j) \sim \frac{f^{\msf{abc}}}{z_{ij}}\sum_{m=0}^{\infty} \frac{1}{m!}  \int_{\mathbb{R}_+} \frac{\d t_i\,t_i^{\Delta_i-\frac{3}{2}+m}}{|1+\frac{\veps_i}{\veps_j}t_i|^{\Delta_i+\Delta_j-1+m}} \left(\frac{\veps_i \bar z_{ij}}{\veps_j} \right)^m \\
\int\,j^{\msf{c}}(\sigma_j)\,\int_{\mathbb{C}^*\times\mathbb{R}_+}\frac{\d s_j}{s_j}\,\,\frac{\d t_j}{t_j^{1-(\Delta_i+\Delta_j-1)}}\,(\im s_j\chi_A(\sigma_j))(\im s_j\chi_B(\sigma_j))\bar\delta^2\left(z_j-s_j\,\lambda(\sigma_j)\right)\,\\
\exp\Big[\im\,t_j\,s_j\mathrm{sgn}(\veps_j+\veps_i\,t_i)\,[\mu(\sigma_j)\,\bar z_j]\Big]\,,
\end{multline}
It is now evident that the last three integrals give us a scalar vertex operator $\Phi^{\msf{c},\mathrm{sgn}(\veps_j+\veps_it_i)}_{\Delta_i+\Delta_j-1, AB}$ at $(z_j,\bar z_j)$. All together, we have a master formula including all $SL(2,\mathbb{R})$ descendants of the OPE with arbitrary orientation configuration:
\begin{multline}
\Gamma^{\msf{a},\veps_i}_{+,\Delta_i,A}(z_i,\bar z_i)\,\Gamma^{\msf{b},\veps_j}_{+,\Delta_j,B}(z_j,\bar z_j) \sim \frac{f^{\msf{abc}}}{z_{ij}}\sum_{m=0}^{\infty} \frac{1}{m!}  \int_{\mathbb{R}_+} \frac{\d t_i\,t_i^{\Delta_i-\frac{3}{2}+m}}{|1+\frac{\veps_i}{\veps_j}t_i|^{\Delta_i+\Delta_j-1+m}}\\ \left(\frac{\veps_i \bar z_{ij}}{\veps_j} \right)^m \bar\partial_j^m\Phi^{\msf{c},\mathrm{sgn}(\veps_j+\veps_it_i)}_{\Delta_i+\Delta_j-1, AB}(z_j,\bar z_j) \,,
\end{multline}
In order to see our favourite Beta functions appearing, first we recall two different integral representations of the Euler Beta function:
\begin{equation}
    B(x,y) = \int_0^{\infty} \frac{t^{x-1}}{(1+t)^{x+y}}\d t = \int_0^1 t^{x-1}(1-t)^{y-1} dt \,,
\end{equation}
Now we start with the case when both gluinos are incoming or outgoing, namely $\veps_i=\veps_j=\veps$, the $t_i$ integral immediately gives 
\begin{multline}
\Gamma^{\msf{a},\veps}_{+,\Delta_i,A}(z_i,\bar z_i)\,\Gamma^{\msf{b},\veps}_{+,\Delta_j,B}(z_j,\bar z_j) \sim \\ \frac{f^{\msf{abc}}}{z_{ij}}\sum_{m=0}^{\infty} \frac{\bar z_{ij}^m}{m!}  B\left(\Delta_i+m-\frac{1}{2},\Delta_j-\frac{1}{2}\right)  \bar\partial_j^m\Phi^{\msf{c},\veps}_{\Delta_i+\Delta_j-1, AB}(z_j,\bar z_j) \,,
\end{multline}
We see that when $m=0$, our coefficient just gives the Beta function in the literature \cite{Fotopoulos:2020bqj}.

Now for the mixed incoming/outgoing case: $\veps_i=-\veps_j=\veps$. We'll need use the alternative expression of the Beta function to split the integral into
\begin{multline}
\frac{f^{\msf{abc}}}{z_{ij}}\,\sum_{m=0}^{\infty}\frac{1}{m!}\int_{0}^{1}\frac{\d t_i\,t_i^{\Delta_i+m-\frac{3}{2}}}{(1-t_i)^{\Delta_i+\Delta_j+m-1}}\,\left(-\bar{z}_{ij}\right)^{m}\,\dbar_j^m\Phi^{\msf{c},-\veps}_{\Delta_i+\Delta_j-1, AB}(z_j,\bar{z}_j) \\
+\frac{f^{\msf{abc}}}{z_{ij}}\,\sum_{m=0}^{\infty}\frac{(-1)^{\Delta_i+\Delta_j}}{m!}\int_{1}^{\infty}\frac{\d t_i\,t_i^{\Delta_i+m-\frac{3}{2}}}{(1-t_i)^{\Delta_i+\Delta_j+m-1}}\,\left(\bar{z}_{ij}\right)^{m}\,\dbar_j^m\Phi^{\msf{c},\veps}_{\Delta_i+\Delta_j-1, AB}(z_j,\bar{z}_j)\,,
\end{multline}
where the first integral just straightforwardly gives 
\begin{multline}
\frac{f^{\msf{abc}}}{z_{ij}}\,\sum_{m=0}^{\infty}\frac{1}{m!}\,B\left(\Delta_i+m-\frac{1}{2},2-\Delta_i-\Delta_j-m\right)\,\left(-\bar{z}_{ij}\right)^{m}\,\dbar_j^m\Phi^{\msf{c},-\veps}_{\Delta_i+\Delta_j-1, AB}(z_j,\bar{z}_j) \,,
\end{multline}
For the second integral, we just need to reparametrize $t_i\mapsto \frac{1}{t_i}$ which gives 
\begin{multline}
\frac{f^{\msf{abc}}}{z_{ij}}\,\sum_{m=0}^{\infty}\frac{(-1)^{\Delta_i+\Delta_j}}{m!}\int_{0}^{1}\frac{\d t_i\,t_i^{\Delta_j-\frac{3}{2}}}{(1-t_i)^{\Delta_i+\Delta_j+m-1}}\,\left(\bar{z}_{ij}\right)^{m}\,\dbar_j^m\Phi^{\msf{c},\veps}_{\Delta_i+\Delta_j-1, AB}(z_j,\bar{z}_j) \\
= \frac{-f^{\msf{abc}}}{z_{ij}}\,\sum_{m=0}^{\infty}\frac{\left(\bar{z}_{ij}\right)^{m}}{m!}\,B\left(\Delta_j-\frac{1}{2},2-\Delta_i-\Delta_j-m\right)\,\dbar_j^m\Phi^{\msf{c},\veps}_{\Delta_i+\Delta_j-1, AB}(z_j,\bar{z}_j) \,,
\end{multline}
where $(-1)^{\Delta_i+\Delta_j}$ flips the orientation of the gluino and gives an overall minus sign.

Combine everything to get 
\begin{multline}
\Gamma^{\msf{a},\veps}_{+,\Delta_i,A}(z_i,\bar z_i)\,\Gamma^{\msf{b},-\veps}_{+,\Delta_j,B}(z_j,\bar z_j) \sim
\frac{-f^{\msf{abc}}}{z_{ij}}\sum_{m=0}^{\infty}\frac{\left(\bar{z}_{ij}\right)^{m}}{m!}\bar\partial_j^m\,\left[B\left(\Delta_j-\frac{1}{2},2-\Delta_i-\Delta_j-m\right)\,\right.\\ 
\left.\Phi^{\msf{c},\veps}_{\Delta_i+\Delta_j-1, AB}(z_j,\bar{z}_j)-(-1)^m B\left(\Delta_i+m-\frac{1}{2},2-\Delta_i-\Delta_j-m\right)\,\Phi^{\msf{c},-\veps}_{\Delta_i+\Delta_j-1, AB}(z_j,\bar{z}_j)  \right],
\end{multline}
Here we see that the mixed orientation OPE coefficients are also contained in the master formula. From now on we shall just write down master formulas containing both orientation configurations and all $SL(2,\mathbb{R})$ descendants for all the other cases, leaving the reader to work out the individual Beta functions.

Now that we have an explicit procedure to compute the like helicity gluino-gluino OPE, we could use this to do other OPEs with slight change in coefficients, for example the like helicity gluon-gluino OPE $\cO^{\msf{a},\veps_i}_{+,\Delta_i}(z_i,\bar z_i)\,\Gamma^{\msf{b},\veps_j}_{+,\Delta_j,A}(z_j,\bar z_j)$: 
\begin{multline}
\cO^{\msf{a},\veps_i}_{+,\Delta_i}(z_i,\bar z_i)\,\Gamma^{\msf{b},\veps_j}_{+,\Delta_j,A}(z_j,\bar z_j) \sim \int\d\sigma_i\,\frac{f^{\msf{abc}}\,j^{\msf{c}}(\sigma_j)}{\sigma_{ij}}\,\frac{\d s_i}{s_i}\,\frac{\d s_j}{s_j}\,\frac{\d t_i}{t_i^{2-\Delta_i}}\,\frac{\d t_j}{t_j^{\frac{3}{2}-\Delta_j}}\,(\im s_j\chi_A(\sigma_j))\, \\
\bar\delta^2\left(z_i-s_i\,\lambda(\sigma_i)\right)\bar\delta^2\left(z_j-s_j\,\lambda(\sigma_j)\right)\,\exp\Big(\im\veps_i\,t_i\,s_i\,[\mu(\sigma_i)\,\bar z_i]+\im\veps_j\,t_j\,s_j\,[\mu(\sigma_j)\,\bar z_j]\Big)\,, 
\end{multline}
Here we shall give the master formula directly after manipulating the integrals following the same procedure as the gluino-gluino case
\begin{multline}
\cO^{\msf{a},\veps_i}_{+,\Delta_i}(z_i,\bar z_i)\,\Gamma^{\msf{b},\veps_j}_{+,\Delta_j,A}(z_j,\bar z_j) \sim  \frac{f^{\msf{abc}}}{z_{ij}}\sum_{m=0}^{\infty} \frac{1}{m!}  \int_{\mathbb{R}_+} \frac{\d t_i\,t_i^{\Delta_i-2+m}}{|1+\frac{\veps_i}{\veps_j}t_i|^{\Delta_i+\Delta_j-\frac{3}{2}+m}} \\
\left(\frac{\veps_i \bar z_{ij}}{\veps_j} \right)^m \bar\partial_j^m\Gamma^{\msf{c},\mathrm{sgn}(\veps_j+\veps_it_i)}_{+,\Delta_i+\Delta_j-1, A}(z_j,\bar z_j) 
\end{multline}
where one could check the coefficients against the literature \cite{Fotopoulos:2020bqj,Jiang:2021xzy} for the $m=0$, $\veps_i=\veps_j=\veps$ case. 

\subsection{Gravitino-gravitino OPE}
Next up we consider the like helicity gravitino-gravitino OPE, the difference compared with the gluino-gluino case is that we no longer have the worldsheet current OPE, instead we have $\mathcal{Z}-\mathcal{W}$ OPEs and $\rho-\tilde\rho$ OPEs given by \eqref{Z-W OPE} and \eqref{rho-rho_OPE}. More specifically, we only use the bosonic part of the $\mathcal{Z}-\mathcal{W}$ OPE: 
\begin{equation}
    \mu^{\Dot{\alpha}}(\sigma_i) \tilde\lambda_{\Dot{\beta}}(\sigma_j) = \frac{\delta^{\Dot{\alpha}}_{\Dot{\beta}}\sqrt{\d\sigma_i\d\sigma_j}}{\sigma_i-\sigma_j}\,,
\end{equation}
First write down the OPE in integral form:
\begin{multline}
\Theta^{\veps_i}_{+,\Delta_i,A}(z_i,\bar z_i)\,\Theta^{\veps_j}_{+,\Delta_j,B}(z_j,\bar z_j)\sim \veps_i\veps_j\int\frac{\d s_i}{s^2_i}\,\frac{\d s_j}{s^2_j}\,\frac{\d t_i}{t_i^{\frac{3}{2}-\Delta_i}}\,\frac{\d t_j}{t_j^{\frac{3}{2}-\Delta_j}}\,\d\sigma_i\,(\im s_i\chi_A(\sigma_i))\,(\im s_j\chi_B(\sigma_j))\\
\,\Big(\im[\tilde\lambda(\sigma_i)\bar z_i]-\veps_is_it_i[\rho(\sigma_i)\bar z_i][\tilde\rho(\sigma_i)\bar z_i] \Big)\,\Big(\im[\tilde\lambda(\sigma_j)\bar z_j]-\veps_js_jt_j[\rho(\sigma_j)\bar z_j][\tilde\rho(\sigma_j)\bar z_j] \Big)\\
\bar\delta^2\left(z_i-s_i\,\lambda(\sigma_i)\right) \,\bar\delta^2\left(z_j-s_j\,\lambda(\sigma_j)\right)\,\exp\Big(\im\veps_i\,t_i\,s_i\,[\mu(\sigma_i)\,\bar z_i]+\im\veps_j\,t_j\,s_j\,[\mu(\sigma_j)\,\bar z_j]\Big)\,,
\end{multline}
Now we take all possible Wick contraction to get a slightly more involved expression:
\begin{align}
    &\Theta^{\veps_i}_{+,\Delta_i,A}(z_i,\bar z_i)\,\Theta^{\veps_j}_{+,\Delta_j,B}(z_j,\bar z_j)\sim \int_{\Sigma_i\times\Sigma_j}  \int_{(\mathbb{C}^*)^2\times(\mathbb{R}_{+})^2}\frac{\d t_i}{t_i^{\frac{3}{2}-\Delta_i}}\frac{\d s_i}{s_i^2}\frac{\d t_j}{t_j^{2-\Delta_i-\Delta_j}}\frac{\d s_j}{s_j^2}\,\veps_i\,\veps_j  \\ \nonumber
    &\,\bar\delta^2\left(z_i-s_i\lambda(\sigma_i)\right)\,\bar\delta^2\left(z_j-s_j\lambda(\sigma_j)\right)\,e^{it_is_it_j\veps_i[\mu(\sigma_i)\bar z_i]+it_js_j\veps_j[\mu(\sigma_j)\bar z_j]}\,(is_i\chi_A(\sigma_i))\,(is_j\chi_B(\sigma_j)) \\ \nonumber
    &\left(\veps_i\veps_jt_it_j^3s_is_j\frac{[\bar z_i\bar z_j]^2}{\sigma_{ij}^2}-\veps_i\veps_jt_it_j^3s_is_j\frac{[\bar z_i\bar z_j]^2}{\sigma_{ij}^2}-i\veps_it_it_j^2s_i\frac{[\bar z_i\bar z_j][\tilde\lambda(\sigma_i)\bar z_i]}{\sigma_{ij}}-i\veps_jt_j^2s_j\frac{[\bar z_j\bar z_i][\tilde\lambda(\sigma_j)\bar z_j]}{\sigma_{ji}} \right. \\ \nonumber
    &\left. +\veps_i^2t_i^2t_j^3s_i^2\frac{[\bar z_i\bar z_j][\tilde\rho(\sigma_i)\bar z_i][\rho(\sigma_i)\bar z_i]}{\sigma_{ij}} +\veps_j^2t_j^3s_j^2\frac{[\bar z_i\bar z_j][\tilde\rho(\sigma_j)\bar z_j][\rho(\sigma_j)\bar z_j]}{\sigma_{ij}}+\veps_j\veps_it_it_j^3s_is_j \right. \\ \nonumber &\left.\frac{[\bar z_i\bar z_j][\tilde\rho(\sigma_i)\bar z_i][\rho(\sigma_j)\bar z_j]}{\sigma_{ij}} +\veps_j\veps_it_it_j^3s_is_j\frac{[\bar z_j\bar z_i][\tilde\rho(\sigma_j)\bar z_j][\rho(\sigma_i)\bar z_i]}{\sigma_{ji}} \right)\,,
\end{align}
where we have rescaled $t_i\mapsto t_it_j$. One notices that double poles appeared but nicely cancel out leaving only simple poles in the OPE, which is rather miraculous in its own right since there were no extra constraints required, the structure of the ambitwistor string vertex operators dictate the simple pole.

Since the structures of the holomorphic functions are exactly the same as in the gluino-gluino case, the expression still localises on $z_{ij}=0$ or $\sigma_{ij}=0$ on the ambitwistor string worldsheet. The roadmap for computing these integrals follows directly from the gluino-gluino case. Hence we perform the $s_i$ integral first against the first holomorphic delta function, integrate by parts to extract residue at $\sigma_{ij}=0$, then use the identities enforced by the remaining delta function to write the pole in a desired form. Since we have already rescaled $t_i\mapsto t_it_j$, we just need to rescale $t_j\mapsto \frac{t_j}{|1+\frac{\veps_i\,t_i}{\veps_j}|}$ and expand around $\bar z_{ij}$ to obtain the entire $SL(2,\mathbb{R})$ tower of all orientation configurations.  

The master formula in this case reads
\begin{multline}
\Theta^{\veps_i}_{+,\Delta_i,A}(z_i,\bar z_i)\,\Theta^{\veps_j}_{+,\Delta_j,B}(z_j,\bar z_j) \sim  \frac{\veps_i\veps_j\bar z_{ij}}{z_{ij}}\sum_{m=0}^{\infty} \frac{1}{m!}  \int_{\mathbb{R}_+} \frac{\d t_i\,t_i^{\Delta_i-\frac{3}{2}+m}}{|1+\frac{\veps_i}{\veps_j}t_i|^{\Delta_i+\Delta_j-1+m}} \\
\left(\frac{\veps_i \bar z_{ij}}{\veps_j} \right)^m \bar\partial_j^m V^{\mathrm{sgn}(\veps_j+\veps_it_i)}_{+,\Delta_i+\Delta_j, AB}(z_j,\bar z_j) \,,
\end{multline}
One could check against the literature for the case with $m=0,\veps_i=\veps_j$, they match up to R-symmetry and the spin $1$ graviphoton $V$, as \cite{Fotopoulos:2020bqj} is considering $\mathcal{N}=1$ case. 

Following very much similar steps, one could go up the ladder in one of the supersymmetric multiplets and compute the like helicity graviton-gravitino OPE:
\begin{multline}
\cG^{\veps_i}_{+,\Delta_i}(z_i,\bar z_i)\,\Theta^{\veps_j}_{+,\Delta_j,A}(z_j,\bar z_j)\sim \veps_i\veps_j\int\frac{\d s_i}{s^2_i}\,\frac{\d s_j}{s^2_j}\,\frac{\d t_i}{t_i^{2-\Delta_i}}\,\frac{\d t_j}{t_j^{\frac{3}{2}-\Delta_j}}\,\d\sigma_i\,\\
(\im s_j\chi_A(\sigma_j))\,\Big(\im[\tilde\lambda(\sigma_i)\bar z_i]-\veps_is_it_i[\rho(\sigma_i)\bar z_i][\tilde\rho(\sigma_i)\bar z_i] \Big)\,\Big(\im[\tilde\lambda(\sigma_j)\bar z_j]-\veps_js_jt_j[\rho(\sigma_j)\bar z_j][\tilde\rho(\sigma_j)\bar z_j] \Big)\\
\bar\delta^2\left(z_i-s_i\,\lambda(\sigma_i)\right) \,\bar\delta^2\left(z_j-s_j\,\lambda(\sigma_j)\right)\,\exp\Big(\im\veps_i\,t_i\,s_i\,[\mu(\sigma_i)\,\bar z_i]+\im\veps_j\,t_j\,s_j\,[\mu(\sigma_j)\,\bar z_j]\Big)\,,
\end{multline}
Here we just give the master formula after evaluating the integrals:
\begin{multline}
\cG^{\veps_i}_{+,\Delta_i}(z_i,\bar z_i)\,\Theta^{\veps_j}_{+,\Delta_j,A}(z_j,\bar z_j) \sim  \frac{\veps_i\veps_j\bar z_{ij}}{z_{ij}}\sum_{m=0}^{\infty} \frac{1}{m!}  \int_{\mathbb{R}_+} \frac{\d t_i\,t_i^{\Delta_i-2+m}}{|1+\frac{\veps_i}{\veps_j}t_i|^{\Delta_i+\Delta_j-\frac{3}{2}+m}} \\
\left(\frac{\veps_i \bar z_{ij}}{\veps_j} \right)^m \bar\partial_j^m\Theta^{\mathrm{sgn}(\veps_j+\veps_it_i)}_{+,\Delta_i+\Delta_j, A}(z_j,\bar z_j) 
\end{multline}

\subsection{Gravitino-gluino OPE}
In this subsection, we consider the mixing OPEs between super Yang-Mills and Einstein super gravity. First  write down the gravitino-gluino OPE: 
\begin{multline}
\Theta^{\veps_i}_{+,\Delta_i,A}(z_i,\bar z_i)\,\Gamma^{\msf{a},\veps_j}_{+,\Delta_j,B}(z_j,\bar z_j)\sim \veps_i\int\frac{\d s_i}{s^2_i}\,\frac{\d s_j}{s_j}\,\frac{\d t_i}{t_i^{\frac{3}{2}-\Delta_i}}\,\frac{\d t_j}{t_j^{\frac{3}{2}-\Delta_j}}\,j^{\msf{a}}(\sigma_j)\,(\im s_i\chi_A(\sigma_i))(\im s_j\chi_B(\sigma_j))\,\\
\Big(\im[\tilde\lambda(\sigma_i)\bar z_i]-\veps_is_it_i[\rho(\sigma_i)\bar z_i][\tilde\rho(\sigma_i)\bar z_i] \Big)\bar\delta^2\left(z_i-s_i\,\lambda(\sigma_i)\right) \,\bar\delta^2\left(z_j-s_j\,\lambda(\sigma_j)\right)\,\\
\exp\Big(\im\veps_i\,t_i\,s_i\,[\mu(\sigma_i)\,\bar z_i]+\im\veps_j\,t_j\,s_j\,[\mu(\sigma_j)\,\bar z_j]\Big)\,,
\end{multline}
Once again, just as the two previous cases, there is no mixing between the two holomorphic delta functions, which suggests that the methodology for this should not differ too much from the other like helicity cases. However, we notice that the OPE we need to consider here is slightly different. The only possible contraction comes from $\im[\tilde\lambda(\sigma_i)\bar z_i]$ and the exponential $e^{i\veps_i\,t_i\,s_i[\mu(\sigma_i)\bar z_i]}$, which just gives us:
\begin{multline}
\Theta^{\veps_i}_{+,\Delta_i,A}(z_i,\bar z_i)\,\Gamma^{\msf{a},\veps_j}_{+,\Delta_j,B}(z_j,\bar z_j)\sim \im \bar z_{ij}\int\frac{\d s_i}{s_i}\,\frac{\d s_j}{s_j}\,\frac{\d t_i}{t_i^{\frac{1}{2}-\Delta_i}}\,\frac{\d t_j}{t_j^{\frac{3}{2}-\Delta_j}}\,\frac{j^{\msf{a}}(\sigma_j)}{\sigma_{ij}}\,(\im s_i\chi_A(\sigma_i))\\
(\im s_j\chi_B(\sigma_j))\,\bar\delta^2\left(z_i-s_i\,\lambda(\sigma_i)\right) \,\bar\delta^2\left(z_j-s_j\,\lambda(\sigma_j)\right)\,\exp\Big(\im\veps_i\,t_i\,s_i\,[\mu(\sigma_i)\,\bar z_i]+\im\veps_j\,t_j\,s_j\,[\mu(\sigma_j)\,\bar z_j]\Big)\,,
\end{multline}
where $\bar z_{ij} :=[\bar z_i\bar z_j]$. We notice that the simple pole emerged in front of the expression, which allows us to integrate by parts to obtain $\bar\delta(\sigma_{ij})$ just as before. Here we simply follow exactly the same steps as in section \ref{same_gluino_gluino} to obtain the final expression containing with all $SL(2,\mathbb{R})$ descendants:
\begin{multline}
\Theta^{\veps_i}_{+,\Delta_i,A}(z_i,\bar z_i)\,\Gamma^{\msf{a},\veps_j}_{+,\Delta_j,B}(z_j,\bar z_j) \sim  \frac{-\veps_i\veps_j\bar z_{ij}}{z_{ij}}\sum_{m=0}^{\infty} \frac{1}{m!}  \int_{\mathbb{R}_+} \frac{\d t_i\,t_i^{\Delta_i-\frac{3}{2}+m}}{|1+\frac{\veps_i}{\veps_j}t_i|^{\Delta_i+\Delta_j+m}} \\
\left(\frac{\veps_i \bar z_{ij}}{\veps_j} \right)^m \bar\partial_j^m\Phi^{\mathrm{sgn}(\veps_j+\veps_it_i)}_{\Delta_i+\Delta_j, AB}(z_j,\bar z_j) \,,
\end{multline}

Similarly, we could go back up the SUSY hierarchy in either multiplets and compute the graviton-gluino OPE and the gravitino-gluon OPE, namely $\cG^{\veps_i}_{+,\Delta_i}(z_i,\bar z_i)\Gamma^{\msf{a},\veps_j}_{+,\Delta_j,A}(z_j,\bar z_j)$ and $\Theta^{\veps_i}_{+,\Delta_i,A}(z_i,\bar z_i)\cO^{\msf{a},\veps_j}_{+,\Delta_j}(z_j,\bar z_j)$. Here we just give their corresponding master formulas.
\begin{multline}
\cG^{\veps_i}_{+,\Delta_i}(z_i,\bar z_i)\Gamma^{\msf{a},\veps_j}_{+,\Delta_j,A}(z_j,\bar z_j) \sim  \frac{\veps_i\veps_j\bar z_{ij}}{z_{ij}}\sum_{m=0}^{\infty} \frac{1}{m!}  \int_{\mathbb{R}_+} \frac{\d t_i\,t_i^{\Delta_i-2+m}}{|1+\frac{\veps_i}{\veps_j}t_i|^{\Delta_i+\Delta_j-\frac{1}{2}+m}} \\
\left(\frac{\veps_i \bar z_{ij}}{\veps_j} \right)^m \bar\partial_j^m\Gamma^{\msf{a},\mathrm{sgn}(\veps_j+\veps_it_i)}_{+,\Delta_i+\Delta_j, A}(z_j,\bar z_j) \,,
\end{multline}
\begin{multline}
\Theta^{\veps_i}_{+,\Delta_i,A}(z_i,\bar z_i)\cO^{\msf{a},\veps_j}_{+,\Delta_j}(z_j,\bar z_j) \sim  \frac{\veps_i\veps_j\bar z_{ij}}{z_{ij}}\sum_{m=0}^{\infty} \frac{1}{m!}  \int_{\mathbb{R}_+} \frac{\d t_i\,t_i^{\Delta_i-\frac{3}{2}+m}}{|1+\frac{\veps_i}{\veps_j}t_i|^{\Delta_i+\Delta_j-\frac{1}{2}+m}} \\
\left(\frac{\veps_i \bar z_{ij}}{\veps_j} \right)^m \bar\partial_j^m\Gamma^{\msf{a},\mathrm{sgn}(\veps_j+\veps_it_i)}_{+,\Delta_i+\Delta_j, A}(z_j,\bar z_j) 
\end{multline}
One can check against the literature \cite{Fotopoulos:2020bqj} for the case $m=0,\veps_i=\veps_j$.

We would like to note here that since the highest possible number of supersymmetry one could have in Einstein-Yang-Mills theories in 4d is 4, we would only be able to compute OPEs between any particle from the gluon multiplet and the graviton, gravitino and the vector in the graviton supermultiplet. Indeed if one attempts to compute, for example, the OPE between a gluon and the scalar in the graviton multiplet, the homogeneity mismatches with the number of supersymmetry in the resulting vertex operator. 


\section{Mixed helicity OPEs}\label{MHO}
In all the like helicity computations, we demonstrated how to extract a single vertex operator out of the integrals and obtain the desired Euler Beta functions by rescaling certain parameters. One would naively expect a similar procedure to work for the mixed helicity OPEs. However, it turns out that the most fundamental observation we made for the like helicity OPEs does not hold anymore, namely the holomorphic delta functions enforcing worldsheet OPEs and celestial OPEs to coincide. The new scattering equations here complexify dramatically, which quickly hinders the steps we developed for the like helicity cases. This is because the new scattering equations now localise the computation in a region in the moduli space where the vertex operators are ill-defined. Hence we would perform certain reparametrization on the affine coordinates $s_i$ and $s_j$ to move to an appropriate patch in the moduli space, where majority of the steps we developed for the like helicity cases would follow through.

Majority of the calculations in this section follow directly from the mixed helicity section of \cite{Adamo:2021zpw}, with an additional subtlety in the fermionic $\chi-\tilde\chi$ OPE. Moreover, we notice that the negative helicity operators we introduced in section \ref{vertex_operator}, the positions of the orientation parameter $\epsilon$ differ from the ones in \cite{Adamo:2021zpw}. In order to avoid the sign ambiguity that could occur, we adopt the more rigorous positioning of $\veps$, which makes slight adaptation to the pure bosonic calculations.

\subsection{Gluino-gluino OPE}\label{mixed-gluino-gluino}
The key difference between mixed helicity OPEs and the like helicity ones is the structure of the holomorphic delta functions. Here we zoom in on the general structure of all mixed helicity OPEs:
\begin{equation}
\bar{\delta}^{2}\!\left(z_i-s_i\,\lambda(\sigma_i)\right) \e^{\im\veps_i\,t_i\,s_i\,[\mu(\sigma_i)\,\bar{z}_i]}\,\bar{\delta}^{2}\!\left(\veps_j\,\bar{z}_j-s_j\,\tilde{\lambda}(\sigma_j)\right)\,\e^{\im\,t_j\,s_j\,\la\tilde{\mu}(\sigma_j)\,z_j\ra}\,,
\end{equation}
First thing we notice is that, due to the $\mathcal{Z}-\mathcal{W}$ OPE \eqref{Z-W OPE}, one could have contractions between $\lambda(\sigma_i)$ in the first delta function and $\tilde\mu(\sigma_j)$ on the exponential. Similarly there could be contractions between $\tilde\lambda(\sigma_j)$ and $\mu(\sigma_i)$. In order to compute such OPEs, we half-Fourier transform the holomorphic delta functions:
\begin{equation}
\begin{split}
\bar{\delta}^{2}\!\left(z_i-s_i\,\lambda(\sigma_i)\right)&=\int_{\C^2}\frac{\d^{2}\msf{m}}{(2\pi)^2}\, \e^{\im\,\la\msf{m}\,z_i\ra-\im\,s_i\,\la\msf{m}\,\lambda(\sigma_i)\ra}\,, \\
\bar{\delta}^{2}\!\left(\veps_j\,\bar{z}_j-s_j\,\tilde{\lambda}(\sigma_j)\right)&=\int_{\C^2}\frac{\d^{2}\tilde{\msf{m}}}{(2\pi)^2}\,\e^{\im\,\veps_j\,[\tilde{\msf{m}}\,\bar{z}_j]-\im\,s_j\,[\tilde{\msf{m}}\,\tilde{\lambda}(\sigma_j)]}\,,
\end{split}
\end{equation}
and use the following rule from Polchinski \cite{Polchinski:1998rq}:
\begin{equation}
\e^{\im\veps_i\,t_i\,s_i\,[\mu(\sigma_i)\,\bar{z}_i]}\,\e^{-\im\,s_j\,[\tilde{\msf{m}}\,\tilde{\lambda}(\sigma_j)]}\sim \exp\left(\im\,\frac{\veps_i\,t_i\,s_i\,s_j}{\sigma_{ij}}\,[\tilde{\msf{m}}\,\bar{z}_i]\right)\,:\e^{\im\veps_i\,t_i\,s_i\,[\mu(\sigma_i)\,\bar{z}_i]}\,\e^{-\im\,s_j\,[\tilde{\msf{m}}\,\tilde{\lambda}(\sigma_j)]}:\,,
\end{equation}
where $:(\cdots):$ indicates normal-ordering.

After properly attending the OPEs above, one could write the mixed helicity gluino-gluino OPE as follows
\begin{multline}
\Gamma^{\msf{a},\veps_i}_{+,\Delta_i,A}(z_i,\bar z_i)\,\bar\Gamma^{\msf{b},\veps_j,B}_{-,\Delta_j}(z_j,\bar z_j) \sim -\int\d\sigma_i\,\frac{\delta_A{}^Bf^{\msf{abc}}\,j^{\msf{c}}(\sigma_j)}{\sigma_{ij}^2}\,\d s_i\,\d s_j\,\frac{\d t_i}{t_i^{\frac{3}{2}-\Delta_i}}\,\frac{\d t_j}{t_j^{\frac{3}{2}-\Delta_j}}\,\\\bar{\delta}^{2}\!\left(z_i-s_{i}\,\lambda(\sigma_i)-\frac{t_j\,s_i\,s_j\,z_j}{\sigma_{ij}}\right)\,
\bar{\delta}^{2}\!\left(\veps_j\,\bar{z}_j-s_j\,\tilde{\lambda}(\sigma_j)+\frac{\veps_i\,t_i\,s_i\,s_j\,\bar{z}_i}{\sigma_{ij}}\right)\,\\
\,\exp\Big(\im\veps_i\,t_i\,s_i\,[\mu(\sigma_i)\,\bar z_i]+\im t_j\,s_j\,\la\tilde\mu(\sigma_j)\, z_j\ra\Big) \,, 
\end{multline}
where we have performed the  $\chi_A(\sigma_i)\tilde\chi^B(\sigma_j)$ OPE sitting in front of the expression according to the fermionic part of the $\mathcal{Z}-\mathcal{W}$ OPE \eqref{Z-W OPE}:
\begin{equation}
    \chi_A(\sigma_i)\tilde\chi^B(\sigma_j) \sim \frac{\delta_A{}^B\,\sqrt{\d\sigma_i\d\sigma_j}}{\sigma_i-\sigma_j}\,,
\end{equation}

At first sight, this expression seems hopeless as we have a double pole $\sigma_{ij}^2$ appearing. However, as we mentioned earlier, this expression is ill defined on the current affine coordinate patch, it requires certain appropriately chosen rescaling to appear regular. Apart from the double pole, we also observe that the third term in each delta function become singular when $\sigma_{ij}\shortrightarrow 0$, unless either $s_i$ or $s_j$ goes to $0$ at the same rate.

Now that we have identified the standing difficulties with the current parametrization of the expression, we notice that a viable rescaling could be either $s_i\mapsto s_i\sigma_{ij}$ or $s_j\mapsto s_j\sigma_{ij}$. These two rescalings will correspond to the holomorphic and anti-holomorphic limits on the celestial sphere respectively, as we mentioned in section \ref{kinematics}. Also notice that either of these rescalings would bring an overall factor of $\sigma_{ij}$, which reduces the power of $\sigma_{ij}$ pole down to $1$. Hence we solved both problems simultaneously using these rescalings. Without loss of generality, we shall begin with the holomorphic rescaling $s_i\mapsto s_i\sigma_{ij}$.
\begin{multline}
\Gamma^{\msf{a},\veps_i}_{+,\Delta_i,A}(z_i,\bar z_i)\,\bar\Gamma^{\msf{b},\veps_j,B}_{-,\Delta_j}(z_j,\bar z_j) \sim -\int\d\sigma_i\,\frac{\delta_A{}^Bf^{\msf{abc}}\,j^{\msf{c}}(\sigma_j)}{\sigma_{ij}}\,\d s_i\,\d s_j\,\frac{\d t_i}{t_i^{\frac{3}{2}-\Delta_i}}\,\frac{\d t_j}{t_j^{\frac{3}{2}-\Delta_j}}\,\\
\bar{\delta}^{2}\!\left(z_i-s_{i}\sigma_{ij}\,\lambda(\sigma_i)-\,t_j\,s_i\,s_j\,z_j\right)\,\bar{\delta}^{2}\!\left(\veps_j\,\bar{z}_j-s_j\,\tilde{\lambda}(\sigma_j)+\veps_i\,t_i\,s_i\,s_j\,\bar{z}_i\right)\,\\ \,\exp\Big(\im\veps_i\,t_i\,s_i\,\sigma_{ij}\,[\mu(\sigma_i)\,\bar z_i]+\im\,t_j\,s_j\,\la\tilde\mu(\sigma_j)\, z_j\ra\Big)\,, 
\end{multline}
Following the roadmap described in the like helicity cases, first we shall perform the $s_i$ integral using the first delta function and get 
\begin{multline}
\Gamma^{\msf{a},\veps_i}_{+,\Delta_i,A}(z_i,\bar z_i)\,\bar\Gamma^{\msf{b},\veps_j,B}_{-,\Delta_j}(z_j,\bar z_j) \sim -\int\d\sigma_i\,\frac{\delta_A{}^Bf^{\msf{abc}}\,j^{\msf{c}}(\sigma_j)}{\sigma_{ij}}\,\d s_j\,\frac{\d t_i}{t_i^{\frac{3}{2}-\Delta_i}}\,\frac{\d t_j}{t_j^{\frac{3}{2}-\Delta_j}}\,\\
\bar{\delta}\!\left(\sigma_{ij}\,\la z_i\lambda(\sigma_i)\ra+t_j\,s_j\,\la z_iz_j\ra\right)\bar{\delta}^{2}\!\left(\veps_j\,\bar{z}_j-s_j\,\tilde{\lambda}(\sigma_j)+\veps_i\,t_i\,s^*_i\,s_j\,\bar{z}_i\right)\, \\ \,\exp\Big(\im\veps_i\,t_i\,s_i^*\,\sigma_{ij}\,[\mu(\sigma_i)\,\bar z_i]+\im t_j\,s_j\,\la\tilde\mu(\sigma_j)\, z_j\ra\Big)\,, 
\end{multline}
where $s^*_i= \frac{1}{\sigma_{ij}\la \iota \lambda(\sigma_i) \ra+t_js_j}$. Now we rescale $t_i\mapsto t_it_j$ and integrate by parts to extract residue at $\sigma_{ij} =0$ 
\begin{multline}
\Gamma^{\msf{a},\veps_i}_{+,\Delta_i,A}(z_i,\bar z_i)\,\bar\Gamma^{\msf{b},\veps_j,B}_{-,\Delta_j}(z_j,\bar z_j) \sim -\int\,\frac{\delta_A{}^Bf^{\msf{abc}}\,j^{\msf{c}}(\sigma_j)}{z_{ij}}\,\frac{\d s_j}{s_j}\,\frac{\d t_i}{t_i^{\frac{3}{2}-\Delta_i}}\,\frac{\d t_j}{t_j^{3-\Delta_i-\Delta_j}}\\
\bar{\delta}^{2}\!\left(\veps_j\,\bar{z}_j-s_j\,\tilde{\lambda}(\sigma_j)+\veps_i\,t_i\,\bar{z}_i\right)\, \,\exp\Big(\im\,t_j\,s_j\,\la\tilde\mu(\sigma_j)\, z_j\ra\Big)\,, 
\end{multline}
Notice that we did not need the identities enforced by the remaining delta function as in the like helicity cases, the $z_{ij}$ pole came out automatically. Another feature of the mixed helicity configuration is that the holomorphic part of the exponential was eliminated straight away, instead the complexity has been shifted to the holomorphic delta function.
Now notice that there is still dependence on $t_i$ inside the remaining delta function, in order to get rid of it, we rewrite the delta function through some algebra
\begin{equation}
    \veps_j\,\bar{z}_j-s_j\,\tilde{\lambda}(\sigma_j)+\veps_i\,t_i\,\bar{z}_i = \veps_j\,(1+\frac{\veps_i}{\veps_j}t_i)\bar z_j-s_j\,\tilde{\lambda}(\sigma_j)+\veps_i\,t_i\,\bar{z}_{ij}\,,
\end{equation}
Now we just need to rescale $s_j\mapsto s_j\,(1+\frac{\veps_i}{\veps_j}t_i)$ to move all $t_i$ dependence inside the delta function to $\bar z_{ij}$ term which we shall Taylor expand around $0$. However, in order to ensure our exponential to be invariant, we also need to rescale $t_j\mapsto \frac{t_j}{|1+\frac{\veps_i}{\veps_j}\,t_i|}$. After these two rescalings we have
\begin{multline}
\Gamma^{\msf{a},\veps_i}_{+,\Delta_i,A}(z_i,\bar z_i)\,\bar\Gamma^{\msf{b},\veps_j,B}_{-,\Delta_j}(z_j,\bar z_j) \sim -\frac{\delta_A{}^Bf^{\msf{abc}}}{z_{ij}}\int_{\mathbb{R}_+}\frac{\d t_i\,t_i^{\Delta_i-\frac{3}{2}}}{|1+\frac{\veps_i}{\veps_j}t_i|^{\Delta_i+\Delta_j}}\int_{\Sigma_j\times\mathbb{C}^*\times\mathbb{R}_+}\,j^{\msf{c}}(\sigma_j)\,\\
\frac{\d s_j}{s_j}\,\frac{\d t_j}{t_j^{3-\Delta_i-\Delta_j}}\bar{\delta}^{2}\!\left(\veps_j\,\bar{z}_j-s_j\,\tilde{\lambda}(\sigma_j)+\frac{\veps_i\veps_j\,t_i}{\veps_j+\veps_i\,t_i}\,\bar{z}_{ij}\right)\, \,\exp\Big(\im\, \mathrm{sgn}(\veps_j+\veps_it_i)\,t_j\,s_j\,\la\tilde\mu(\sigma_j)\, z_j\ra\Big)\,, 
\end{multline}
where we recognize the last three integrals represent a negative helicity gluon vertex operator. In the end to get all the $SL(2,\mathbb{R})$ descendants to appear in our expression, we simply need to Taylor expand the remaining delta function around $\bar z_{ij} =0$ to obtain the master formula in the mixed helicity case:
\begin{multline}
\Gamma^{\msf{a},\veps_i}_{+,\Delta_i,A}(z_i,\bar z_i)\,\bar\Gamma^{\msf{b},\veps_j,B}_{-,\Delta_j}(z_j,\bar z_j) \sim  \frac{\delta_A{}^Bf^{\msf{abc}}}{z_{ij}}\sum_{m=0}^{\infty} \frac{1}{m!}  \int_{\mathbb{R}_+} \frac{\d t_i\,t_i^{\Delta_i-\frac{3}{2}+m}}{|1+\frac{\veps_i}{\veps_j}t_i|^{\Delta_i+\Delta_j}} \\
\left(\frac{\veps_i\veps_j \bar z_{ij}}{\veps_j+\veps_i\,t_i} \right)^m \bar\partial_j^m\cO^{\msf{c},\mathrm{sgn}(\veps_j+\veps_it_i)}_{-,\Delta_i+\Delta_j-1}(z_j,\bar z_j) \,,
\end{multline}
Similarly, if we rescale $s_j\mapsto s_j\sigma_{ij}$, we would obtain a positive helicity gluon in the end. Combining the holomorphic part and the anti-holomorphic part we have 
\begin{multline}
\Gamma^{\msf{a},\veps_i}_{+,\Delta_i,A}(z_i,\bar z_i)\,\bar\Gamma^{\msf{b},\veps_j,B}_{-,\Delta_j}(z_j,\bar z_j) \sim  \frac{\delta_A{}^Bf^{\msf{abc}}}{z_{ij}}\sum_{m=0}^{\infty} \frac{1}{m!}  \int_{\mathbb{R}_+} \frac{\d t_i\,t_i^{\Delta_i-\frac{3}{2}+m}}{|1+\frac{\veps_i}{\veps_j}t_i|^{\Delta_i+\Delta_j}} \\
\left(\frac{\veps_i\veps_j \bar z_{ij}}{\veps_j+\veps_i\,t_i} \right)^m \bar\partial_j^m\cO^{\msf{c},\mathrm{sgn}(\veps_j+\veps_it_i)}_{-,\Delta_i+\Delta_j-1}(z_j,\bar z_j) \\
+\frac{\delta_A{}^Bf^{\msf{abc}}}{\bar z_{ji}}\sum_{m=0}^{\infty} \frac{1}{m!}  \int_{\mathbb{R}_+} \frac{\d t_j\,t_j^{\Delta_j-\frac{3}{2}+m}}{|1+\frac{\veps_j}{\veps_i}t_j|^{\Delta_i+\Delta_j}} 
\left(\frac{\veps_j\veps_i z_{ji}}{\veps_i+\veps_j\,t_j} \right)^m \partial_i^m\cO^{\msf{c},\mathrm{sgn}(\veps_i+\veps_jt_j)}_{+,\Delta_i+\Delta_j-1}(z_i,\bar z_i) \,,
\end{multline}
which essentially flips holomorphicity and interchanges $i$ and $j$.

Using practically the same procedure, we could also compute the mixed helicity gluon-gluino OPE:
\begin{multline}
\cO^{\msf{a},\veps_i}_{+,\Delta_i}(z_i,\bar z_i)\,\bar\Gamma^{\msf{b},\veps_j,A}_{-,\Delta_j}(z_j,\bar z_j) \sim \int\d\sigma_i\,\frac{f^{\msf{abc}}\,j^{\msf{c}}(\sigma_j)}{\sigma_{ij}}\,\frac{\d s_i}{s_i}\,\frac{\d s_j}{s_j}\,\frac{\d t_i}{t_i^{2-\Delta_i}}\,\frac{\d t_j}{t_j^{\frac{3}{2}-\Delta_j}}\,(\im s_j\tilde\chi^A(\sigma_j))\\
\bar{\delta}^{2}\!\left(z_i-s_{i}\,\lambda(\sigma_i)-\frac{t_j\,s_i\,s_j\,z_j}{\sigma_{ij}}\right)\bar{\delta}^{2}\!\left(\veps_j\,\bar{z}_j-s_j\,\tilde{\lambda}(\sigma_j)+\frac{\veps_i\,t_i\,s_i\,s_j\,\bar{z}_i}{\sigma_{ij}}\right)\, \\ \,\exp\Big(\im\veps_i\,t_i\,s_i\,[\mu(\sigma_i)\,\bar z_i]+\im\,t_j\,s_j\,\la\tilde\mu(\sigma_j)\, z_j\ra\Big)\,, 
\end{multline}
Note that in this case we could only rescale $s_i$ as rescaling $s_j$ cancels the $\sigma_{ij}$ pole and the OPE becomes non-singular. Here we write down the master formula obtained 
\begin{multline}
\cO^{\msf{a},\veps_i}_{+,\Delta_i}(z_i,\bar z_i)\,\bar\Gamma^{\msf{b},\veps_j,A}_{-,\Delta_j}(z_j,\bar z_j) \sim  \frac{f^{\msf{abc}}}{z_{ij}}\sum_{m=0}^{\infty} \frac{1}{m!}  \int_{\mathbb{R}_+} \frac{\d t_i\,t_i^{\Delta_i-2+m}}{|1+\frac{\veps_i}{\veps_j}t_i|^{-\frac{1}{2}+\Delta_i+\Delta_j}} \\
\left(\frac{\veps_i\veps_j \bar z_{ij}}{\veps_j+\veps_i\,t_i} \right)^m \bar\partial_j^m\bar\Gamma^{\msf{c},\mathrm{sgn}(\veps_j+\veps_it_i),A}_{-,\Delta_i+\Delta_j-1}(z_j,\bar z_j) 
\end{multline}
One could check against the literature \cite{Fotopoulos:2020bqj,Jiang:2021xzy} for the case $m=0,\veps_i=\veps_j$.


\subsection{Gravitino-gravitino OPE}
Now we would like to compute the mixed helicity gravitino-gravitino OPE, after treating all the $\mathcal{Z}-\mathcal{W}$ OPEs, we have the same holomorphic delta functions as before:
\begin{multline}
\Theta^{\veps_i}_{+,\Delta_i,A}(z_i,\bar z_i)\,\bar\Theta^{\veps_j,B}_{-,\Delta_j}(z_j,\bar z_j)\sim \veps_i\int\frac{\d s_i}{s^2_i}\,\frac{\d s_j}{s^2_j}\,\frac{\d t_i}{t_i^{\frac{3}{2}-\Delta_i}}\,\frac{\d t_j}{t_j^{\frac{3}{2}-\Delta_j}}\,\d\sigma_i\,\Big(\im[\tilde\lambda(\sigma_i)\bar z_i]-\veps_is_it_i[\rho(\sigma_i)\bar z_i][\tilde\rho(\sigma_i)\bar z_i] \Big)\\
\Big(\im\la\lambda(\sigma_j)\,z_j\ra-s_jt_j\la\tilde\rho(\sigma_j) \,z_j\ra\la\rho(\sigma_j)\,z_j\ra \Big)\,(\im s_i\chi_A(\sigma_i))\,(\im s_j\tilde\chi^B(\sigma_j))\,\bar{\delta}^{2}\!\left(z_i-s_{i}\,\lambda(\sigma_i)-\frac{t_j\,s_i\,s_j\,z_j}{\sigma_{ij}}\right)\\
\bar{\delta}^{2}\!\left(\veps_j\,\bar{z}_j-s_j\,\tilde{\lambda}(\sigma_j)+\frac{\veps_i\,t_i\,s_i\,s_j\,\bar{z}_i}{\sigma_{ij}}\right)\,  \exp\Big(\im\veps_i\,t_i\,s_i\,[\mu(\sigma_i)\,\bar z_i]+\im\,t_j\,s_j\,\la\tilde\mu(\sigma_j)\, z_j\ra\Big)\,, 
\end{multline}
However, notice that when we perform the $\chi-\tilde\chi$ OPE $(\im s_i\chi_A(\sigma_i))\,(\im s_j\tilde\chi^B(\sigma_j))$, we will have a simple $\sigma_{ij}$ pole, following the strategy developed in section \ref{mixed-gluino-gluino}, we need to rescale either $s_i\mapsto s_i\sigma_{ij}$ or $s_j\mapsto s_j\sigma_{ij}$ to make the holomorphic delta function appear normal as in the like helicity cases again. However, to keep the simple pole in front, we need to ensure that the rescaling does not change the power of our pole, after combining all $s_i$ and $s_j$ in our expression, we see that the $s$ integrals read $\int_{(\mathbb{C}^*)^2}\frac{\d s_i}{s_i}\frac{\d s_j}{s_j}$, which certainly does not affect the pole when rescaled. From here we simply follow the computation in section \ref{mixed-gluino-gluino}, which should give us the holomorphic part and the anti-holomorphic part at the same time when we choose to rescale $s_i$ or $s_j$. 

The master formula we obtain for this can be written as 
\begin{multline}\label{mixed_gravitino_gravitino_OPE}
\Theta^{\veps_i}_{+,\Delta_i,A}(z_i,\bar z_i)\,\bar\Theta^{\veps_j,B}_{-,\Delta_j}(z_j,\bar z_j) \sim  \frac{\im\veps_i\delta_A{}^B\,\bar z_{ij}}{z_{ij}}\sum_{m=0}^{\infty} \frac{1}{m!}  \int_{\mathbb{R}_+} \frac{\d t_i\,t_i^{\Delta_i-\frac{3}{2}+m}}{|1+\frac{\veps_i}{\veps_j}t_i|^{2+\Delta_i+\Delta_j}} \\
\left(\frac{\veps_i\veps_j \bar z_{ij}}{\veps_j+\veps_i\,t_i} \right)^m \bar\partial_j^m\cG^{\mathrm{sgn}(\veps_j+\veps_it_i)}_{-,\Delta_i+\Delta_j}(z_j,\bar z_j) \\
+\frac{\im\veps_i\delta_A{}^B\,z_{ji}}{\veps_j\,\bar z_{ji}}\sum_{m=0}^{\infty} \frac{1}{m!}  \int_{\mathbb{R}_+} \frac{\d t_j\,t_j^{\Delta_j-\frac{3}{2}+m}}{|1+\frac{\veps_j}{\veps_i}t_j|^{2+\Delta_i+\Delta_j}} 
\left(\frac{\veps_j\veps_i  z_{ji}}{\veps_i+\veps_j\,t_j} \right)^m \partial_i^m\cG^{\mathrm{sgn}(\veps_i+\veps_jt_j)}_{+,\Delta_j+\Delta_i}(z_i,\bar z_i) \,,
\end{multline}
Analogously, we could also compute the graviton-gravitino mixed helicity OPE:
\begin{multline}
\cG^{\veps_i}_{+,\Delta_i}(z_i,\bar z_i)\,\bar\Theta^{\veps_j,A}_{-,\Delta_j}(z_j,\bar z_j)\sim \veps_i\int\frac{\d s_i}{s^2_i}\,\frac{\d s_j}{s^2_j}\,\frac{\d t_i}{t_i^{2-\Delta_i}}\,\frac{\d t_j}{t_j^{\frac{3}{2}-\Delta_j}}\,\d\sigma_i\,\Big(\im[\tilde\lambda(\sigma_i)\bar z_i]-\veps_is_it_i[\rho(\sigma_i)\bar z_i][\tilde\rho(\sigma_i)\bar z_i] \Big)\\
\Big(\im\la\lambda(\sigma_j)\,z_j\ra-s_jt_j\la\tilde\rho(\sigma_j) \,z_j\ra\la\rho(\sigma_j)\,z_j\ra \Big)\,(\im s_j\tilde\chi^A(\sigma_j))\,\bar{\delta}^{2}\!\left(z_i-s_{i}\,\lambda(\sigma_i)-\frac{t_j\,s_i\,s_j\,z_j}{\sigma_{ij}}\right)\\
\bar{\delta}^{2}\!\left(\veps_j\,\bar{z}_j-s_j\,\tilde{\lambda}(\sigma_j)+\frac{\veps_i\,t_i\,s_i\,s_j\,\bar{z}_i}{\sigma_{ij}}\right)\,  \exp\Big(\im\veps_i\,t_i\,s_i\,[\mu(\sigma_i)\,\bar z_i]+\im\,t_j\,s_j\,\la\tilde\mu(\sigma_j)\, z_j\ra\Big)\,, 
\end{multline}
Notice that there is no longer any $\sigma_{ij}$ pole generated by the $\chi-\tilde\chi$ OPE, however, when we attempts to regularise the holomorphic delta functions by rescaling $s_i\mapsto s_i\sigma_{ij}$, we see that the homogeneity of $s_i$ here generates a  $\frac{1}{\sigma_{ij}}$ pole for us for free. 
Note that one could only rescale $s_i$ here since rescaling $s_j$ would lead to us having a non-singular OPE. Here we just give the resulting master formula:
\begin{multline}
\cG^{\veps_i}_{+,\Delta_i}(z_i,\bar z_i)\,\bar\Theta^{\veps_j,A}_{-,\Delta_j}(z_j,\bar z_j)\sim  \frac{\im\veps_i\,\bar z_{ij}}{z_{ij}}\sum_{m=0}^{\infty} \frac{1}{m!}  \int_{\mathbb{R}_+} \frac{\d t_i\,t_i^{\Delta_i-2+m}}{|1+\frac{\veps_i}{\veps_j}t_i|^{\frac{3}{2}+\Delta_i+\Delta_j}} \\
\left(\frac{\veps_i\veps_j \bar z_{ij}}{\veps_j+\veps_i\,t_i} \right)^m \bar\partial_j^m\bar\Theta^{\mathrm{sgn}(\veps_j+\veps_it_i),A}_{-,\Delta_i+\Delta_j}(z_j,\bar z_j) 
\end{multline}
One could check against the literature \cite{Fotopoulos:2020bqj} for the case $m=0,\veps_i=\veps_j$ except for the R-symmetry factor.

\subsection{Gravitino-gluino OPE}
Now we have both mixed helicity OPEs for SYM and SUGRA, just as in the like helicity section, we would like to compute the mixed helicity EYM OPEs. We begin with the gravitino-gluino mixed helicity OPE:
\begin{multline}
\Theta^{\veps_i}_{+,\Delta_i,A}(z_i,\bar z_i)\,\bar\Gamma^{\msf{a},\veps_j,B}_{-,\Delta_j}(z_j,\bar z_j)\sim \veps_i\int\frac{\d s_i}{s^2_i}\,\frac{\d s_j}{s_j}\,\frac{\d t_i}{t_i^{\frac{3}{2}-\Delta_i}}\,\frac{\d t_j}{t_j^{\frac{3}{2}-\Delta_j}}\,j^{\msf{a}}(\sigma_j)\,\d\sigma_i\,\\
(\im s_i\chi_A(\sigma_i))(\im s_j\tilde\chi^B(\sigma_j))\,\Big(\im[\tilde\lambda(\sigma_i)\bar z_i]-\veps_is_it_i[\rho(\sigma_i)\bar z_i][\tilde\rho(\sigma_i)\bar z_i] \Big)\,\bar{\delta}^{2}\!\left(z_i-s_{i}\,\lambda(\sigma_i)-\frac{t_j\,s_i\,s_j\,z_j}{\sigma_{ij}}\right)\\
\bar{\delta}^{2}\!\left(\veps_j\,\bar{z}_j-s_j\,\tilde{\lambda}(\sigma_j)+\frac{\veps_i\,t_i\,s_i\,s_j\,\bar{z}_i}{\sigma_{ij}}\right)\,  \exp\Big(\im\veps_i\,t_i\,s_i\,[\mu(\sigma_i)\,\bar z_i]+\im\,t_j\,s_j\,\la\tilde\mu(\sigma_j)\, z_j\ra\Big)\,, 
\end{multline}
Here we have the $\chi-\tilde\chi$ OPE giving us a simple pole, however, we notice that we could only rescale $s_i$ here to have maintain the simple $\sigma_{ij}$ pole. Hence one only ends up with the anti-holomorphic part, to get the holomorphic part describing a positive helicity gluon, we will need the opposite helicity configuration $\bar\Theta^{\veps_i,A}_{-,\Delta_i}(z_i,\bar z_i)\,\Gamma^{\msf{a},\veps_j}_{+,\Delta_j,B}(z_j,\bar z_j)$.
The master formula we obtain is
\begin{multline}
\Theta^{\veps_i}_{+,\Delta_i,A}(z_i,\bar z_i)\,\bar\Gamma^{\msf{a},\veps_j,B}_{-,\Delta_j}(z_j,\bar z_j)\sim  \frac{-\im\veps_i\,\delta_A{}^B\,\bar z_{ij}}{z_{ij}}\sum_{m=0}^{\infty} \frac{1}{m!}  \int_{\mathbb{R}_+} \frac{\d t_i\,t_i^{\Delta_i-\frac{3}{2}+m}}{|1+\frac{\veps_i}{\veps_j}t_i|^{1+\Delta_i+\Delta_j}} \\
\left(\frac{\veps_i\veps_j \bar z_{ij}}{\veps_j+\veps_i\,t_i} \right)^m \bar\partial_j^m\cO^{\msf{a},\mathrm{sgn}(\veps_j+\veps_it_i)}_{-,\Delta_i+\Delta_j}(z_j,\bar z_j) \,,
\end{multline}
Now we could follow the same steps to compute two other OPEs, namely the graviton-gluino OPE and the gravitino-gluon OPE. Firstly the graviton-gluino OPE:
\begin{multline}
\cG^{\veps_i}_{+,\Delta_i}(z_i,\bar z_i)\,\bar\Gamma^{\msf{a},\veps_j,A}_{-,\Delta_j}(z_j,\bar z_j)\sim \veps_i\int\frac{\d s_i}{s^2_i}\,\frac{\d s_j}{s_j}\,\frac{\d t_i}{t_i^{2-\Delta_i}}\,\frac{\d t_j}{t_j^{\frac{3}{2}-\Delta_j}}\,j^{\msf{a}}(\sigma_j)\,\d\sigma_i\,\\
(\im s_j\tilde\chi^A(\sigma_j))\,\Big(\im[\tilde\lambda(\sigma_i)\bar z_i]-\veps_is_it_i[\rho(\sigma_i)\bar z_i][\tilde\rho(\sigma_i)\bar z_i] \Big)\,\bar{\delta}^{2}\!\left(z_i-s_{i}\,\lambda(\sigma_i)-\frac{t_j\,s_i\,s_j\,z_j}{\sigma_{ij}}\right)\\
\bar{\delta}^{2}\!\left(\veps_j\,\bar{z}_j-s_j\,\tilde{\lambda}(\sigma_j)+\frac{\veps_i\,t_i\,s_i\,s_j\,\bar{z}_i}{\sigma_{ij}}\right)\,  \exp\Big(\im\veps_i\,t_i\,s_i\,[\mu(\sigma_i)\,\bar z_i]+\im\,t_j\,s_j\,\la\tilde\mu(\sigma_j)\, z_j\ra\Big)\,, 
\end{multline}
Observe that we could only rescale $s_i$ here as rescaling $s_j$ gives non-singular OPE.
The master formula can be written as:
\begin{multline}
\cG^{\veps_i}_{+,\Delta_i}(z_i,\bar z_i)\,\bar\Gamma^{\msf{a},\veps_j,A}_{-,\Delta_j}(z_j,\bar z_j)\sim  \frac{\im\veps_i\,\bar z_{ij}}{z_{ij}}\sum_{m=0}^{\infty} \frac{1}{m!}  \int_{\mathbb{R}_+} \frac{\d t_i\,t_i^{\Delta_i-2+m}}{|1+\frac{\veps_i}{\veps_j}t_i|^{\frac{1}{2}+\Delta_i+\Delta_j}} \\
\left(\frac{\veps_i\veps_j \bar z_{ij}}{\veps_j+\veps_i\,t_i} \right)^m \bar\partial_j^m\bar\Gamma^{\msf{a},\mathrm{sgn}(\veps_j+\veps_it_i)}_{-,\Delta_i+\Delta_j}(z_j,\bar z_j) \,,
\end{multline}
With the gravitino-gluon OPE:
\begin{multline}
\Theta^{\veps_i}_{+,\Delta_i,A}(z_i,\bar z_i)\,\cO^{\msf{a},\veps_j}_{-,\Delta_j}(z_j,\bar z_j)\sim \veps_i\int\frac{\d s_i}{s^2_i}\,\frac{\d s_j}{s_j}\,\frac{\d t_i}{t_i^{\frac{3}{2}-\Delta_i}}\,\frac{\d t_j}{t_j^{2-\Delta_j}}\,j^{\msf{a}}(\sigma_j)\,\d\sigma_i\,\\
(\im s_i\chi_A(\sigma_i))\,\Big(\im[\tilde\lambda(\sigma_i)\bar z_i]-\veps_is_it_i[\rho(\sigma_i)\bar z_i][\tilde\rho(\sigma_i)\bar z_i] \Big)\,\bar{\delta}^{2}\!\left(z_i-s_{i}\,\lambda(\sigma_i)-\frac{t_j\,s_i\,s_j\,z_j}{\sigma_{ij}}\right)\\
\bar{\delta}^{2}\!\left(\veps_j\,\bar{z}_j-s_j\,\tilde{\lambda}(\sigma_j)+\frac{\veps_i\,t_i\,s_i\,s_j\,\bar{z}_i}{\sigma_{ij}}\right)\,  \exp\Big(\im\veps_i\,t_i\,s_i\,[\mu(\sigma_i)\,\bar z_i]+\im\,t_j\,s_j\,\la\tilde\mu(\sigma_j)\, z_j\ra\Big) 
\end{multline}
Notice that no matter we rescale $s_i$ or $s_j$, the expression refuses to give us any $\sigma_{ij}$ pole. This means that the OPE between a gravitino and a gluon is always regular, which agrees with the statement in the literature \cite{Fotopoulos:2020bqj}. This comes from the fact that there is no Lorentz invariant 3-vertex for such configuration. This can also be checked by BCFW methods in \cite{Himwich:2021dau}.

\subsection{Scalar OPE}
Here we address the discussion we had in section \ref{set_up} about splitting our spectrum between twistor space and dual twistor space. We mentioned that the only ambiguity is with the scalars in $\mathcal{N}=4$ SYM and $\mathcal{N}=8$ SUGRA, where half of the scalars originate from the positive helicity multiplet and half from the negative one. We shall see that both of the representations are needed to compute the scalar-scalar OPEs. To keep things simple and illustrate our point, we demonstrate the calculation for the gluon scalars. First take two scalars originated from the positive helicity multiplet:
\begin{multline}
\Phi^{\msf{a},\veps_i}_{\Delta_i,AB}(z_i,\bar z_i)\,\Phi^{\msf{b},\veps_j}_{\Delta_j,CD}(z_j,\bar z_j) \sim \int\d\sigma_i\,\frac{f^{\msf{abc}}\,j^{\msf{c}}(\sigma_j)}{\sigma_{ij}}\,\frac{\d s_i}{s_i}\,\frac{\d s_j}{s_j}\,\frac{\d t_i}{t_i^{1-\Delta_i}}\,\frac{\d t_j}{t_j^{1-\Delta_j}}\,\\
(-s_i^2\chi_A(\sigma_i)\chi_B(\sigma_i))(- s_j^2\chi_C(\sigma_j)\chi_D(\sigma_j))\,\bar{\delta}^{2}\!\left(z_i-s_{i}\,\lambda(\sigma_i)\right)\,\\
\bar{\delta}^{2}\!\left(z_j-s_j\,\lambda(\sigma_j)\right)\,\exp\Big(\im\veps_i\,t_i\,s_i\,[\mu(\sigma_i)\,\bar z_i]+\im \veps_j\, t_j\,s_j\,[\mu(\sigma_j)\,\bar z_j]\Big)\,,
\end{multline}
From the number of fermionic indices on the right hand side, it is straightforward to deduce the resulting vertex operator should carry $4$ R-symmetry indices with homogeneity $-4$ on twistor space. However, there is no such particle present in the spectrum. If one were to ignore this and proceed with the computation naively as in section \ref{same_gluino_gluino}, one would end up with
\begin{multline}
\Phi^{\msf{a},\veps_i}_{\Delta_i,AB}(z_i,\bar z_i)\,\Phi^{\msf{b},\veps_j}_{\Delta_j,CD}(z_j,\bar z_j) \sim \int \, \frac{f^{\msf{a}\msf{b}\msf{c}}j^{\msf{c}}(\sigma_j)}{z_{ij}}\, \d s_j\, s_j \,\frac{\d t_i}{t_i^{1-\Delta_i}}\,\frac{\d t_j}{t_j^{1-\Delta_j}}\,\\ (\chi_A(\sigma_j)\chi_B(\sigma_j)\chi_C(\sigma_j)\chi_D(\sigma_j))\,\bar\delta^2(z_j-s_j\lambda(\sigma_j))\,\exp\Big(\im\veps_i\,t_i\,s_j\,[\mu(\sigma_j)\,\bar z_i]+\im \veps_j\, t_j\,s_j\,[\mu(\sigma_j)\,\bar z_j]\Big) \,,
\end{multline}
where the resulting expression has homogeneity $-2$ on twistor space, which disagrees with the number of $\chi$ it contains. As the OPE between two scalars of negative helicity origin is very much similar, we proceed to consider the remaining option, namely the scalar-scalar OPE with opposite helicity origin:
\begin{multline}
\Phi^{\msf{a},\veps_i}_{\Delta_i,AB}(z_i,\bar z_i)\,\Phi^{\msf{b},\veps_j,CD}_{\Delta_j}(z_j,\bar z_j) \sim \int\,\frac{f^{\msf{abc}}\,j^{\msf{c}}(\sigma_j)}{\sigma_{ij}}\,\frac{\d s_i}{s_i}\,\frac{\d s_j}{s_j}\,\frac{\d t_i}{t_i^{1-\Delta_i}}\,\frac{\d t_j}{t_j^{1-\Delta_j}}\,\\
(-s_i^2\chi_A(\sigma_i)\chi_B(\sigma_i))(- s_j^2\tilde\chi^C(\sigma_j)\tilde\chi^D(\sigma_j))\,\bar{\delta}^{2}\!\left(z_i-s_{i}\,\lambda(\sigma_i)-\frac{t_j\,s_i\,s_j\,z_j}{\sigma_{ij}}\right)\,\\
\bar{\delta}^{2}\!\left(\veps_j\,\bar{z}_j-s_j\,\tilde{\lambda}(\sigma_j)+\frac{\veps_i\,t_i\,s_i\,s_j\,\bar{z}_i}{\sigma_{ij}}\right)\,\exp\Big(\im\veps_i\,t_i\,s_i\,[\mu(\sigma_i)\,\bar z_i]+\im t_j\,s_j\,\la\tilde\mu(\sigma_j)\, z_j\ra\Big)\,,
\end{multline}
First the order of $\sigma_{ij}$ pole we have after computing the $\chi-\tilde\chi$ OPEs is $3$, however notice that by rescaling $s_i\mapsto s_i\sigma_{ij}$ or $s_j\mapsto s_j\sigma_{ij}$, the pole becomes $\frac{1}{\sigma_{ij}}$. Proceed with rescaling $s_i\mapsto s_i\sigma_{ij}$ and we see that after performing the $s_i$ integral and integrating by parts:
\begin{multline}
\Phi^{\msf{a},\veps_i}_{\Delta_i,AB}(z_i,\bar z_i)\,\Phi^{\msf{b},\veps_j,CD}_{\Delta_j}(z_j,\bar z_j) \sim \int \, \frac{f^{\msf{a}\msf{b}\msf{c}}j^{\msf{c}}(\sigma_j)\,\veps_{AB}{}^{CD}}{z_{ij}}\, \frac{\d s_j}{s_j} \,\frac{\d t_i}{t_i^{1-\Delta_i}}\,\frac{\d t_j}{t_j^{3-\Delta_j}}\,\\ \bar\delta^2\left(\veps_j\,\bar z_j-s_j\tilde\lambda(\sigma_j)+\frac{\veps_i\,t_i\,\bar z_i}{t_j}\right)\,\exp\Big(\im \, t_j\,s_j\,\la\tilde\mu(\sigma_j)\,z_j\ra\Big)
\end{multline}
where the $s_i$ integral provided us with the number of $s_j$ needed to obtain homogeneity $0$, which agrees perfectly with the number of $\chi$ remaining. Indeed we see that a negative helicity gluon vertex operator is a few steps away from the expression on the right hand side. If we were to proceed with rescaling on $s_j$ instead of $s_i$, a positive helicity gluon vertex operator will appear on the right hand side. 
The OPE for the $\mathcal{N}=8$ scalar-scalar is very much similar, we summarize the master formulas in equation \eqref{A.14} and \eqref{A.32} in the appendix.

\section{Supersymmetric Holographic symmetry}\label{SSS}
Following the steps in section 5 in \cite{Adamo:2021zpw}, one could obtain the soft algebra of gluino-gluon, gravitino-graviton, graviton-gluino and gravitino-gluon by shifting and relabeling the indices. We shall see that the algebras remain invariant as the purely non-supersymmetric cases, which agrees with recent discovery in the literature \cite{Ahn:2021erj}. 

\subsection{SYM soft symmetries}
Since we have explicit representations of the vertex operators of all particle content in our framework, to see soft symmetries, one just needs to take residues at certain values of the conformal scaling dimension $\Delta$ and then perform the OPE. For demonstration purposes, we start with the like helicity gluon-gluino both outgoing scenario. It turns out that the most convenient way of expressing our vertex operator here is the integrated form of \eqref{sym_multiplet_+}:  
\begin{equation}
    \Gamma^{\msf{a}}_{+,\Delta,A}(z,\bar z) = \im\int j^\msf{a}(\sigma)\,\chi_A(\sigma)\,\frac{\la \iota\lambda(\sigma)\ra^{\Delta-\frac{1}{2}}}{\la \iota z\ra^{\Delta-\frac{1}{2}}}\,\bar\delta\left(\la\lambda(\sigma)\,z\ra\right)\,\frac{(-\im)^{\frac{1}{2}-\Delta}\,\Gamma(\Delta-\frac{1}{2})}{[\mu(\sigma)\,\bar z]^{\Delta-\frac{1}{2}}}\,,
\end{equation}
where the $s$-integral and the $t$-integral have been performed. The soft gluinos are defined to be the residues at half integer values $\Delta= k+\frac{1}{2}$, where $k\in\{0, -1,-2, ...\}$. 
\begin{multline}
    L^{\msf{a}}_{+,k+\frac{1}{2},A}(z,\bar z) :=\text{Res}_{\Delta=k+\frac{1}{2}}\,\Gamma^{\msf{a}}_{+,\Delta,A}(z,\bar z) \\
    = \frac{1}{2\pi}\oint \frac{(-\im)^{-k}\,j^\msf{a}(\sigma)}{(-k)!}\frac{[\mu(\sigma)\bar z]^{-k}\la\iota\lambda(\sigma)\ra^{k}}{\la\lambda(\sigma)z\ra}\chi_A(\sigma)\,,
\end{multline}
Notice that the soft gluino vertex operator here is similar compared to the soft gluon vertex operator, hence by relabeling $k=3-2p$ and binomial expand $[\mu(\sigma)\bar z]= \mu^{\Dot{0}}+\bar z\mu^{\Dot{1}}$ in $\bar z$:
\begin{equation}
    L^{\msf{a}}_{+,\frac{7}{2}-2p,A}(z,\bar z) 
    = \sum_{m=\frac{3}{2}-p}^{p-\frac{3}{2}}\frac{\bar z^{p-m-\frac{3}{2}}S^{\msf{a},p}_{m,A}(z)}{\Gamma(p-m-\frac{1}{2})\,\Gamma(p+m-\frac{1}{2})}\,,
\end{equation}
where $g^p_m(\sigma)=(\mu^{\Dot{0}})^{p+m-\frac{3}{2}}(\mu^{\Dot{1}})^{p-m-\frac{3}{2}}$, $p$ runs from $\frac{3}{2},2,\frac{5}{2},...$. And 
\begin{equation}
    S^{\msf{a},p}_{m,A}(z)= \frac{\im^{2p-2}}{2\pi\im}\oint\frac{j^{\msf{a}}(\sigma)\,g^p_m(\sigma)}{\la\iota\lambda(\sigma)\ra^{2p-3}\la\lambda(\sigma)z\ra}\chi_A(\sigma) \,,
\end{equation}
Here we also take the expression for soft gluons from \cite{Adamo:2021zpw}:
\begin{equation}
    R^{\msf{a}}_{+,3-2q}(z,\bar z) 
    = \sum_{n=1-q}^{q-1}\frac{\bar z^{q-n-1}S^{\msf{a},q}_{n}(z)}{\Gamma(q-n)\,\Gamma(q+n)}\,,
\end{equation}
where
\begin{equation}
    S^{\msf{a},q}_{n}(z) = \frac{\im^{2q-2}}{2\pi\im}\oint\frac{j^{\msf{a}}(\sigma)\,\tilde g^q_n(\sigma)}{\la\iota\lambda(\sigma)\ra^{2q-3}\la\lambda(\sigma)z\ra} \,,
\end{equation}
with $\tilde g^q_n(\sigma)=(\mu^{\Dot{0}})^{q+n-1}(\mu^{\Dot{1}})^{q-n-1}$ and $q$ runs from $1,\frac{3}{2},2,...$. Using techniques to compute like helicity gluino-gluino OPEs in section \ref{same_gluino_gluino}, we have for gluon-gluino:
\begin{equation}
    S^{\msf{a},p}_{m}(z_i)\,S^{\msf{b},q}_{n,A}(z_j) \sim \frac{f^{\msf{a}\msf{b}\msf{c}}}{z_{ij}}\, S^{\msf{c},p+q-1}_{m+n,A}(z_j) 
\end{equation}
\subsection{SUGRA soft symmetries}
Next up we consider the like helicity outgoing-outgoing graviton-gravitino soft OPE, to do this, first we write the outgoing positive helicity gravitino vertex operator in the following integrated form:
\begin{multline}
    \Theta_{+,\Delta,A}(z,\bar z) = \im\int \Big(\tilde\lambda^{\Dot{\alpha}}\frac{\partial}{\partial\mu^{\Dot{\alpha}}} +\tilde\rho^{\Dot{\alpha}}\rho^{\Dot{\beta}}\frac{\partial^2}{\partial\mu^{\Dot{\alpha}}\partial\mu^{\Dot{\beta}}} \Big)\,\chi_A(\sigma)\,\frac{\la \iota\lambda(\sigma)\ra^{\Delta+\frac{1}{2}}}{\la \iota z\ra^{\Delta+\frac{1}{2}}}\\
    \,\bar\delta\left(\la\lambda(\sigma)\,z\ra\right)\,\frac{(-\im)^{-\frac{1}{2}-\Delta}\,\Gamma(\Delta-\frac{3}{2})}{[\mu(\sigma)\,\bar z]^{\Delta-\frac{3}{2}}}\,,
\end{multline}
The soft gravitinos $I_{+,k+\frac{1}{2},A}(z,\bar z)$ are defined to be the residues at half integer values $\Delta = k+\frac{1}{2}$, where $k\in\{1,0,-1,...\}$. Although this is not the conventional way to label the indices, it is of importance during the computation, we shall see that we could unwind such strange labeling towards the end of the calculation.
\begin{multline}
    I_{+,k+\frac{1}{2},A}(z,\bar z) := \text{Res}_{\Delta= k+\frac{1}{2}}\Theta_{+,\Delta,A}(z,\bar z)\\
    = \frac{1}{2\pi}\oint\Big(\tilde\lambda^{\Dot{\alpha}}\frac{\partial}{\partial\mu^{\Dot{\alpha}}} +\tilde\rho^{\Dot{\alpha}}\rho^{\Dot{\beta}}\frac{\partial^2}{\partial\mu^{\Dot{\alpha}}\partial\mu^{\Dot{\beta}}} \Big)\,\frac{\im^{-1-k}[\mu(\sigma)\bar z]^{1-k}\la\iota\lambda(\sigma)\ra^{k+1}}{(1-k)!\la\lambda(\sigma)z\ra}\chi_A(\sigma)\,,
\end{multline}
This is similar compared to the soft graviton vertex operator excluding supersymmetry and weight in $\lambda(\sigma)$. In order to expand $I_{+,k+\frac{1}{2},A}(z,\bar z)$ in soft modes, we relabel $k= 4-2p$ and binomial expand $[\mu(\sigma)\bar z]= \mu^{\Dot{0}}+\bar z\mu^{\Dot{1}}$ in $\bar z$:
\begin{equation}
    I_{+,\frac{9}{2}-2p,A}(z,\bar z)
    = \sum_{m=\frac{3}{2}-p}^{p-\frac{3}{2}}\frac{\bar z^{p-m-\frac{3}{2}}\,w^{p}_{m,A}(z)}{\Gamma(p-m-\frac{1}{2})\,\Gamma(p+m-\frac{1}{2})}\,,
\end{equation}
where $g^p_m(\sigma)=(\mu^{\Dot{0}})^{p+m-\frac{3}{2}}(\mu^{\Dot{1}})^{p-m-\frac{3}{2}}$ just as in the SYM case with $p\in\{\frac{3}{2},2,\frac{5}{2}...\}$, and the soft modes $w^{p}_{m,A}(z)$ are defined as:
\begin{equation}
    w^{p}_{m,A}(z) = 
    \frac{\im^{2p}}{2\pi\im}\oint\Big(\tilde\lambda^{\Dot{\alpha}}\frac{\partial g^p_m(\sigma)}{\partial\mu^{\Dot{\alpha}}} +\tilde\rho^{\Dot{\alpha}}\rho^{\Dot{\beta}}\frac{\partial^2 g^p_m(\sigma)}{\partial\mu^{\Dot{\alpha}}\partial\mu^{\Dot{\beta}}} \Big)
    \frac{\chi_A(\sigma)}{\la\iota\lambda(\sigma)\ra^{2p-5}\la\lambda(\sigma)z\ra}\,,
\end{equation}
Together with soft graviton modes from \cite{Adamo:2021zpw}:
\begin{equation}
    w^{q}_{n}(z) = 
    \frac{\im^{2q}}{2\pi\im}\oint\Big(\tilde\lambda^{\Dot{\alpha}}\frac{\partial\tilde g^q_n(\sigma)}{\partial\mu^{\Dot{\alpha}}} +\tilde\rho^{\Dot{\alpha}}\rho^{\Dot{\beta}}\frac{\partial^2\tilde g^q_n(\sigma)}{\partial\mu^{\Dot{\alpha}}\partial\mu^{\Dot{\beta}}} \Big)
    \frac{1}{\la\iota\lambda(\sigma)\ra^{2q-5}\la\lambda(\sigma)z\ra}\,,
\end{equation}
where $\tilde g^q_n(\sigma)=(\mu^{\Dot{0}})^{q+n-1}(\mu^{\Dot{1}})^{q-n-1}$ and $q$ runs from $1,\frac{3}{2},2,...$. we obtain the soft graviton-gravitino OPE:
\begin{equation}
    w^p_{m,A}(z_i)w^q_{n}(z_j) \sim
    \frac{2\left(m\left(q-1\right)-n\left(p-\frac{3}{2}\right)\right)}{z_{ij}}\,w^{p+q-2}_{m+n,A}(z_j)\,,
\end{equation}
where we have used the fact that:
\begin{equation}
    \left\{g^p_m,\tilde g^q_n\right\} := \veps^{\dal\Dot{\beta}}\frac{\partial g^p_m}{\partial \mu^{\dal}}\frac{\partial \tilde g^q_n}{\partial \mu^{\Dot{\beta}}} = 2\left(m\left(q-1\right)-n\left(p-\frac{3}{2}\right)\right)g^{p+q-2}_{m+n}\,,
\end{equation}
Because of the convention we have chosen, the index $p$ begins at $\frac{3}{2}$ instead of $1$, to make algebra look just as the usual infinite dimensional symmetry algebra introduced in \cite{Strominger:2021lvk}, we simply relabel $p$ by $p+\frac{1}{2}$. The algebra we obtain is:
\begin{equation}
    w^p_{m,A}(z_i)w^q_{n}(z_j) \sim \frac{2(m(q-1)-n(p-1))}{z_{ij}}\,w^{p+q-2}_{m+n,A}(z_j)\,,
\end{equation}
Notice that the soft expansion and binomial expansion we consider here do not differ from the pure bosonic case, only carrying an extra factor of $\chi_A(\sigma)$ which is not present in the bosonic case. Hence the $w$ algebra here is still the diffeomorphism of the $\mu^{\dal}$ plane. However, one could consider doing the soft and binomial expansion on the entire super multiplet and then take the OPE, in which case the fermionic coordinate $\chi_A(\sigma)$ on twistor space will also need to be expanded. Then we see that the algebra we obtain is a SUSY extension of the diffeomorphism of the $\mu^{\dal}$ plane, the diffeomorphism of the $\mu^{\dal}-\chi$ hypersurface.

It is straightforward to consider the supersymmetric soft Einstein-Yang-Mills OPEs, namely graviton-gluino and gluon-gravitino. To do this we simply take binomial expansions of the corresponding soft particles and take their OPE. We still consider the outgoing-outgoing like helicity configuration. Here we just present the results, which stay invariant as the purely bosonic soft gluon-graviton algebra:
\begin{equation}
    w^p_m(z_i)S^q_{n,A}(z_j) \sim \frac{2(m(q-1)-n(p-1))}{z_{ij}}\,S^{p+q-2}_{m+n,A}(z_j)\,,
\end{equation}
for the graviton-gluino and 
\begin{equation}
    w^p_{m,A}(z_i)S^q_{n}(z_j) \sim \frac{2(m(q-1)-n(p-1))}{z_{ij}}\,S^{p+q-2}_{m+n,A}(z_j)
\end{equation}
for the gravitino-gluon.

\subsection{Soft-hard OPE}
So far we have seen hard-hard and soft-soft OPEs in maximally supersymmetric theories, it is worthwhile to consider is the action of a soft particle acting on a hard one. One essentially just follows the steps in section 5 in \cite{Adamo:2021zpw}. Here we simply state the results. 
A soft gluino acting on a hard gluon yields:
\begin{multline}
    L^{\msf{a}}_{+,k-\frac{1}{2},A}(z_i,\bar z_i)\, \cO^{\msf{b},\veps}_{J,\Delta}(z_j,\bar z_j) \sim\\ \frac{\veps^{k-1}}{(1-k)!}\,\frac{f^{\msf{a}\msf{b}\msf{c}}}{z_{ij}}\,
    \prod_{r=1}^{1-k}\left(\bar z_{ij}\,\bar\partial_j-2\bar h+r\right)\,\Gamma^{\msf{c},\veps}_{J-\mathrm{sgn}(J)/2,\Delta+k-\frac{3}{2},A}(z_j,\bar z_j)\,,
\end{multline}
where $k\in\{1,0,-1,...\}$, $J=\pm 1$ denotes helicity of the gluon, $\mathrm{sgn}(J)=\pm 1$ is the sign of $J$ and $\bar h=(\Delta-J)/2$. 

If the order is reversed, a soft gluon acting a hard gluino gives us:
\begin{equation}
    R^{\msf{a}}_{+,k}(z_i,\bar z_i)\,
    \Gamma^{\msf{b},\veps}_{J,\Delta,A}(z_j,\bar z_j) \sim \frac{\veps^{k-1}}{(1-k)!}\,\frac{f^{\msf{a}\msf{b}\msf{c}}}{z_{ij}}\,\prod_{r=1}^{1-k}\left(\bar z_{ij}\,\bar\partial_j-2\bar h+r\right)\,\Gamma^{\msf{c},\veps}_{J,\Delta+k-1,A}(z_j,\bar z_j)\,,
\end{equation}
where now $k\in\{1,0,-1,...\}$, $J=\pm \frac{1}{2}$ denotes the helicity of the gluino and $\bar h=(\Delta-J)/2$.

For the gravitino-graviton soft-hard OPE, we have for a soft gravitino acting on a hard graviton:
\begin{multline}
   I_{+,k-\frac{1}{2},A}(z_i,\bar z_i)\, \cG^{\veps}_{J,\Delta}(z_j,\bar z_j) \sim \\ \frac{-\veps^{k}}{(1-k)!}\,\frac{\bar z_{ij}}{z_{ij}}\,
   \prod_{r=1}^{1-k}\left(\bar z_{ij}\,\bar\partial_j-2\bar h-1+r\right)\,\Theta^{\veps}_{J-\mathrm{sgn}(J)/2,\Delta+k-\frac{1}{2},A}(z_j,\bar z_j)\,,
\end{multline}
where $k\in\{2,1,0,...\}$, $J=\pm 2$ denotes the helicity of the graviton and $\bar h = (\Delta-J)/2$.

The OPE with the reversed order, namely a soft graviton acting on a hard gravitino reads:
\begin{equation}
    H_{+,k}(z_i,\bar z_i)\,\Theta^{\veps}_{J,\Delta,A}(z_j,\bar z_j)\sim \frac{-\veps^{k}}{(1-k)!}\,\frac{\bar z_{ij}}{z_{ij}}\,\prod_{r=1}^{1-k}\left(\bar z_{ij}\,\bar\partial_j-2\bar h-1+r\right)\,\Theta^{\veps}_{J,\Delta+k,A}(z_j,\bar z_j)\,,
\end{equation}
where $k\in\{2,1,0,...\}$, $J=\pm \frac{3}{2}$ now denotes the helicity of the gravitino and $\bar h = (\Delta-J)/2$.

Similarly, we could consider the action of a soft gravitino on a hard gluon:
\begin{multline}
    I_{+,k-\frac{1}{2},A}(z_i,\bar z_i)\,\cO^{\msf{a},\veps}_{J,\Delta}(z_j,\bar z_j) \sim\\ \frac{-\veps^{k}}{(1-k)!}\,\frac{\bar z_{ij}}{z_{ij}}\,
    \prod_{r=1}^{1-k}\left(\bar z_{ij}\,\bar\partial_j-2\bar h-1+r\right)\,\Gamma^{\msf{a},\veps}_{J-\mathrm{sgn}(J)/2,\Delta+k-\frac{1}{2},A}(z_j,\bar z_j)\,,
\end{multline}
where $k\in\{2,1,0,...\}$, $J=\pm 1$ denotes the helicity of the gluon and $\bar h = (\Delta-J)/2$.

The other super EYM OPE we could consider is a soft graviton acting on a hard gluino:
\begin{equation}
    H_{+,k}(z_i,\bar z_i)\,\Gamma^{\msf{a},\veps}_{J,\Delta,A}(z_j,\bar z_j) \sim \frac{-\veps^{k}}{(1-k)!}\,\frac{\bar z_{ij}}{z_{ij}}\,\prod_{r=1}^{1-k}\left(\bar z_{ij}\,\bar\partial_j-2\bar h-1+r\right)\,\Gamma^{\msf{a},\veps}_{J,\Delta+k,A}(z_j,\bar z_j)\,,
\end{equation}
where $k\in\{2,1,0,...\}$, $J=\pm \frac{1}{2}$ denotes the helicity of the gluino and $\bar h = (\Delta-J)/2$. 

One could also expand the product acting on the vertex operators into a sum, where it essentially follows the identity:
\begin{multline}
   \frac{-\veps^{k}}{(1-k)!}\,\prod_{r=1}^{1-k}\left(\bar z_{ij}\,\bar\partial_j-2\bar h-1+r\right)\,\cU_A(z_j,\bar z_j) = \\ \sum_{l=0}^{1-k}\frac{(-1)^{k+l}\,\veps^{k}}{l!\,(1-k-l)!}\frac{\Gamma(2\bar h+1)}{\Gamma(2\bar h+k+l)}\,\bar z_{ij}^l\,\bar\partial_j^l\,\cU_A(z_j,\bar z_j)\,,
\end{multline}
\begin{multline}
   \frac{\veps^{k-1}}{(1-k)!}\,\prod_{r=1}^{1-k}\left(\bar z_{ij}\,\bar\partial_j-2\bar h+r\right)\,\cU_A(z_j,\bar z_j) = \\ \sum_{l=0}^{1-k}\frac{(-1)^{k+l-1}\,\veps^{k-1}}{l!\,(1-k-l)!}\frac{\Gamma(2\bar h)}{\Gamma(2\bar h+k+l-1)}\,\bar z_{ij}^l\,\bar\partial_j^l\,\cU_A(z_j,\bar z_j)
\end{multline}
where $\cU_A$ denotes either a gluino or a gravitino. Substituting these identities in our expressions, we see that they match results from \cite{Jiang:2021ovh,Himwich:2021dau}.

\acknowledgments
I would like to thank Yvonne Geyer for conversations and sharing her notes on 4d ambitwistor strings, and Tim Adamo, Eduardo Casali and Atul Sharma for conversations and inspirations for this project and commenting on the script. I would also like to thank an anonymous referee for thorough review of the manuscript which helped greatly in improving the draft. WB is supported by a Royal Society PhD Studentship. 

\appendix
\section{Summary of all OPEs}

From here on we list all singular collinear OPEs computable within our framework in the form of master equations. The roadmap from the master equation to obtain the Euler Beta functions for different orientation configurations is laid out in section \ref{same_gluino_gluino}. The literature on this \cite{Fotopoulos:2020bqj,Jiang:2021xzy} have focused on the leading order and incoming-incoming or outgoing-outgoing configuration where $\veps_i=\veps_j$. One could easily extract the Euler Beta function coefficients from the following master formulas and check against the existing ones in the literature, where we see that they match up to $R$ symmetry and scalars. For the rest of the OPEs, to the best of our knowledge, we believe this is the first time they have been written down, which we list here for future references.  

\subsection{$\mathcal{N}=4$ SYM}
\begin{align}
     \Gamma^{\msf{a},\veps_i}_{+,\Delta_i,A}\,\Gamma^{\msf{b},\veps_j}_{+,\Delta_j,B} \sim & \frac{f^{\msf{abc}}}{z_{ij}}\sum_{m=0}^{\infty} \frac{1}{m!}  \int_{\mathbb{R}_+} \frac{\d t_i\,t_i^{\Delta_i-\frac{3}{2}+m}}{|1+\frac{\veps_i}{\veps_j}t_i|^{\Delta_i+\Delta_j-1+m}} \label{A.1} \\ &\left(\frac{\veps_i \bar z_{ij}}{\veps_j} \right)^m \bar\partial_j^m\Phi^{\msf{c},\mathrm{sgn}(\veps_j+\veps_it_i)}_{\Delta_i+\Delta_j-1, AB}(z_j,\bar z_j) \nonumber \\
     \cO^{\msf{a},\veps_i}_{+,\Delta_i}(z_i,\bar z_i)\,\Gamma^{\msf{b},\veps_j}_{+,\Delta_j,A}(z_j,\bar z_j) \sim & \frac{f^{\msf{abc}}}{z_{ij}}\sum_{m=0}^{\infty} \frac{1}{m!}  \int_{\mathbb{R}_+} \frac{\d t_i\,t_i^{\Delta_i-2+m}}{|1+\frac{\veps_i}{\veps_j}t_i|^{\Delta_i+\Delta_j-\frac{3}{2}+m}} \\
     &\left(\frac{\veps_i \bar z_{ij}}{\veps_j} \right)^m \bar\partial_j^m\Gamma^{\msf{c},\mathrm{sgn}(\veps_j+\veps_it_i)}_{+,\Delta_i+\Delta_j-1, A}(z_j,\bar z_j) \nonumber \\
     \Gamma^{\msf{a},\veps_i}_{+,\Delta_i,A}(z_i,\bar z_i)\,\bar\Gamma^{\msf{b},\veps_j,B}_{-,\Delta_j}(z_j,\bar z_j) \sim & \frac{\delta_A{}^Bf^{\msf{abc}}}{z_{ij}}\sum_{m=0}^{\infty} \frac{1}{m!}  \int_{\mathbb{R}_+} \frac{\d t_i\,t_i^{\Delta_i-\frac{3}{2}+m}}{|1+\frac{\veps_i}{\veps_j}t_i|^{\Delta_i+\Delta_j}} \\
     &\left(\frac{\veps_i\veps_j \bar z_{ij}}{\veps_j+\veps_i\,t_i} \right)^m \bar\partial_j^m\cO^{\msf{c},\mathrm{sgn}(\veps_j+\veps_it_i)}_{-,\Delta_i+\Delta_j-1}(z_j,\bar z_j)\nonumber \\
     +\frac{\delta_A{}^B\,f^{\msf{abc}}}{\bar z_{ji}}\sum_{m=0}^{\infty} \frac{1}{m!}  \int_{\mathbb{R}_+} &\frac{\d t_j\,t_j^{\Delta_j-\frac{3}{2}+m}}{|1+\frac{\veps_j}{\veps_i}t_j|^{\Delta_i+\Delta_j}} 
    \left(\frac{\veps_j\veps_i z_{ji}}{\veps_i+\veps_j\,t_j} \right)^m \partial_i^m\cO^{\msf{c},\mathrm{sgn}(\veps_i+\veps_jt_j)}_{+,\Delta_i+\Delta_j-1}(z_i,\bar z_i) \nonumber \\
    \cO^{\msf{a},\veps_i}_{+,\Delta_i}(z_i,\bar z_i)\,\bar\Gamma^{\msf{b},\veps_j,A}_{-,\Delta_j}(z_j,\bar z_j) \sim & \frac{ f^{\msf{abc}}}{z_{ij}}\sum_{m=0}^{\infty} \frac{1}{m!}  \int_{\mathbb{R}_+} \frac{\d t_i\,t_i^{\Delta_i-2+m}}{|1+\frac{\veps_i}{\veps_j}t_i|^{-\frac{1}{2}+\Delta_i+\Delta_j}}\\
    &\left(\frac{\veps_i\veps_j \bar z_{ij}}{\veps_j+\veps_i\,t_i} \right)^m \bar\partial_j^m\bar\Gamma^{\msf{c},\mathrm{sgn}(\veps_j+\veps_it_i),A}_{-,\Delta_i+\Delta_j-1}(z_j,\bar z_j)\nonumber 
\end{align}
\begin{align}
    \Phi^{\msf{a},\veps_i}_{\Delta_i,AB}(z_i,\bar z_i)\,\Phi^{\msf{b},\veps_j,CD}_{\Delta_j}(z_j,\bar z_j)\sim & \frac{\veps_{AB}{}^{CD}\,f^{\msf{abc}}}{z_{ij}}\sum_{m=0}^{\infty} \frac{1}{m!}  \int_{\mathbb{R}_+} \frac{\d t_i\,t_i^{\Delta_i-1+m}}{|1+\frac{\veps_i}{\veps_j}t_i|^{\Delta_i+\Delta_j}} \label{A.14} \\
    &\left(\frac{\veps_i\veps_j \bar z_{ij}}{\veps_j+\veps_i\,t_i} \right)^m \bar\partial_j^m\cO^{\msf{c},\mathrm{sgn}(\veps_j+\veps_it_i)}_{-,\Delta_i+\Delta_j-1}(z_j,\bar z_j) \nonumber \\
    +\frac{\veps_{AB}{}^{CD}\,f^{\msf{abc}}}{\bar z_{ji}}\sum_{m=0}^{\infty} \frac{1}{m!}  \int_{\mathbb{R}_+} &\frac{\d t_j\,t_j^{\Delta_j-1+m}}{|1+\frac{\veps_j}{\veps_i}t_j|^{\Delta_i+\Delta_j}} 
    \left(\frac{\veps_j\veps_i z_{ji}}{\veps_i+\veps_j\,t_j} \right)^m  \partial_i^m\cO^{\msf{c},\mathrm{sgn}(\veps_i+\veps_jt_j)}_{+,\Delta_i+\Delta_j-1}(z_i,\bar z_i) \nonumber \\
    \cO^{\msf{a},\veps_i}_{+,\Delta_i}(z_i,\bar z_i) \Phi^{\msf{b},\veps_j,AB}_{\Delta_j}(z_j,\bar z_j) \sim & \frac{f^{\msf{abc}}}{z_{ij}}\sum_{m=0}^{\infty} \frac{1}{m!}  \int_{\mathbb{R}_+} \frac{\d t_i\,t_i^{\Delta_i-2+m}}{|1+\frac{\veps_i}{\veps_j}t_i|^{\Delta_i+\Delta_j-1}} \\ &\left(\frac{\veps_i\veps_j \bar z_{ij}}{\veps_j+\veps_i\,t_i} \right)^m \bar\partial_j^m\Phi^{\msf{c},\mathrm{sgn}(\veps_j+\veps_it_i),AB}_{\Delta_i+\Delta_j-1}(z_j,\bar z_j) \nonumber \\
    \Gamma^{\msf{a},\veps_i}_{+,\Delta_i,A}(z_i,\bar z_i) \Phi^{\msf{b},\veps_j,BC}_{\Delta_j}(z_j,\bar z_j) \sim & \frac{\delta_A{}^{[B} f^{\msf{a}\msf{b}\msf{c}}}{z_{ij}}\sum_{m=0}^{\infty} \frac{1}{m!}  \int_{\mathbb{R}_+} \frac{\d t_i\,t_i^{\Delta_i-\frac{3}{2}+m}}{|1+\frac{\veps_i}{\veps_j}t_i|^{\Delta_i+\Delta_j-\frac{1}{2}}} \\ & \left(\frac{\veps_i\veps_j \bar z_{ij}}{\veps_j+\veps_i\,t_i} \right)^m \bar\partial_j^m\bar\Gamma^{\msf{c},\mathrm{sgn}(\veps_j+\veps_it_i),C]}_{-,\Delta_i+\Delta_j-1}(z_j,\bar z_j) \nonumber 
\end{align}
\subsection{$\mathcal{N}=8$ SUGRA}
\begin{align}
    \Theta^{\veps_i}_{+,\Delta_i,A}(z_i,\bar z_i)\,\Theta^{\veps_j}_{+,\Delta_j,B}(z_j,\bar z_j) \sim & \frac{\veps_i\veps_j\bar z_{ij}}{z_{ij}}\sum_{m=0}^{\infty} \frac{1}{m!}  \int_{\mathbb{R}_+} \frac{\d t_i\,t_i^{\Delta_i-\frac{3}{2}+m}}{|1+\frac{\veps_i}{\veps_j}t_i|^{\Delta_i+\Delta_j-1+m}} \\
     &\left(\frac{\veps_i \bar z_{ij}}{\veps_j} \right)^m \bar\partial_j^m V^{\mathrm{sgn}(\veps_j+\veps_it_i)}_{+,\Delta_i+\Delta_j, AB}(z_j,\bar z_j) \nonumber \\
     \cG^{\veps_i}_{+,\Delta_i}(z_i,\bar z_i)\,\Theta^{\veps_j}_{+,\Delta_j,A}(z_j,\bar z_j) \sim & \frac{\veps_i\veps_j\bar z_{ij}}{z_{ij}}\sum_{m=0}^{\infty} \frac{1}{m!}  \int_{\mathbb{R}_+} \frac{\d t_i\,t_i^{\Delta_i-2+m}}{|1+\frac{\veps_i}{\veps_j}t_i|^{\Delta_i+\Delta_j-\frac{3}{2}+m}} \\
    &\left(\frac{\veps_i \bar z_{ij}}{\veps_j} \right)^m \bar\partial_j^m\Theta^{\mathrm{sgn}(\veps_j+\veps_it_i)}_{+,\Delta_i+\Delta_j, A}(z_j,\bar z_j) \nonumber \\
    \Theta^{\veps_i}_{+,\Delta_i,A}(z_i,\bar z_i)\,\bar\Theta^{\veps_j,B}_{-,\Delta_j}(z_j,\bar z_j) \sim & \frac{\im\veps_i\delta_A{}^B\,\bar z_{ij}}{z_{ij}}\sum_{m=0}^{\infty} \frac{1}{m!}  \int_{\mathbb{R}_+} \frac{\d t_i\,t_i^{\Delta_i-\frac{3}{2}+m}}{|1+\frac{\veps_i}{\veps_j}t_i|^{2+\Delta_i+\Delta_j}} \\
    &\left(\frac{\veps_i\veps_j \bar z_{ij}}{\veps_j+\veps_i\,t_i} \right)^m \bar\partial_j^m\cG^{\mathrm{sgn}(\veps_j+\veps_it_i)}_{-,\Delta_i+\Delta_j}(z_j,\bar z_j) \nonumber \\
    +\frac{\im\veps_i\delta_A{}^B\,z_{ji}}{\veps_j\,\bar z_{ji}}\sum_{m=0}^{\infty} \frac{1}{m!}  \int_{\mathbb{R}_+}& \frac{\d t_j\,t_j^{\Delta_j-\frac{3}{2}+m}}{|1+\frac{\veps_j}{\veps_i}t_j|^{2+\Delta_i+\Delta_j}} 
    \left(\frac{\veps_j\veps_i  z_{ji}}{\veps_i+\veps_j\,t_j} \right)^m \partial_i^m\cG^{\mathrm{sgn}(\veps_i+\veps_jt_j)}_{+,\Delta_j+\Delta_i}(z_i,\bar z_i) \nonumber \\
    \cG^{\veps_i}_{+,\Delta_i}(z_i,\bar z_i)\,\bar\Theta^{\veps_j,A}_{-,\Delta_j}(z_j,\bar z_j)\sim & \frac{\im\veps_i\,\bar z_{ij}}{z_{ij}}\sum_{m=0}^{\infty} \frac{1}{m!}  \int_{\mathbb{R}_+} \frac{\d t_i\,t_i^{\Delta_i-2+m}}{|1+\frac{\veps_i}{\veps_j}t_i|^{\frac{3}{2}+\Delta_i+\Delta_j}} \\
    &\left(\frac{\veps_i\veps_j \bar z_{ij}}{\veps_j+\veps_i\,t_i} \right)^m \bar\partial_j^m\bar\Theta^{\mathrm{sgn}(\veps_j+\veps_it_i),A}_{-,\Delta_i+\Delta_j}(z_j,\bar z_j) \nonumber \\
    \cG^{\veps_i}_{+,\Delta_i}(z_i,\bar z_i) \bar V^{\veps_j,AB}_{-,\Delta_j}(z_j,\bar z_j) \sim & \frac{\im\,\veps_i\,\bar z_{ij} }{z_{ij}}\sum_{m=0}^{\infty} \frac{1}{m!}  \int_{\mathbb{R}_+} \frac{\d t_i\,t_i^{\Delta_i-2+m}}{|1+\frac{\veps_i}{\veps_j}t_i|^{\Delta_i+\Delta_j+1}} \\ & \left(\frac{\veps_i\veps_j \bar z_{ij}}{\veps_j+\veps_i\,t_i} \right)^m \bar\partial_j^m \bar V^{\mathrm{sgn}(\veps_j+\veps_it_i),AB}_{-,\Delta_i+\Delta_j}(z_j,\bar z_j) \nonumber \\
    \cG^{\veps_i}_{+,\Delta_i}(z_i,\bar z_i)  V^{\veps_j}_{+,\Delta_j,AB}(z_j,\bar z_j) \sim & \frac{\veps_i\,\veps_j\,\bar z_{ij} }{z_{ij}}\sum_{m=0}^{\infty} \frac{1}{m!}  \int_{\mathbb{R}_+} \frac{\d t_i\,t_i^{\Delta_i-2+m}}{|1+\frac{\veps_i}{\veps_j}t_i|^{\Delta_i+\Delta_j-1}} \\ & \left(\frac{\veps_i \bar z_{ij}}{\veps_j} \right)^m \bar\partial_j^m V^{\mathrm{sgn}(\veps_j+\veps_it_i)}_{+,\Delta_i+\Delta_j,AB}(z_j,\bar z_j) \nonumber \\
    \Theta^{\veps_i}_{+,\Delta_i,A}(z_i,\bar z_i) \bar V^{\veps_j,BC}_{-,\Delta_j}(z_j,\bar z_j) \sim & \frac{-\delta_A{}^{[B}\,\veps_i\,\bar z_{ij} }{z_{ij}}\sum_{m=0}^{\infty} \frac{1}{m!}  \int_{\mathbb{R}_+} \frac{\d t_i\,t_i^{\Delta_i-\frac{3}{2}+m}}{|1+\frac{\veps_i}{\veps_j}t_i|^{\Delta_i+\Delta_j+\frac{3}{2}}} \\ & \left(\frac{\veps_i\veps_j \bar z_{ij}}{\veps_j+\veps_i\,t_i} \right)^m \bar\partial_j^m\bar\Theta^{\mathrm{sgn}(\veps_j+\veps_it_i),C]}_{-,\Delta_i+\Delta_j}(z_j,\bar z_j) \nonumber \\
    \Theta^{\veps_i}_{+,\Delta_i,A}(z_i,\bar z_i) V^{\veps_j}_{+,\Delta_j,BC}(z_j,\bar z_j) \sim & \frac{\veps_i\veps_j\,\bar z_{ij} }{z_{ij}}\sum_{m=0}^{\infty} \frac{1}{m!}  \int_{\mathbb{R}_+} \frac{\d t_i\,t_i^{\Delta_i-\frac{3}{2}+m}}{|1+\frac{\veps_i}{\veps_j}t_i|^{\Delta_i+\Delta_j-\frac{1}{2}}} \\ & \left(\frac{\veps_i \bar z_{ij}}{\veps_j} \right)^m \bar\partial_j^m\Xi^{\mathrm{sgn}(\veps_j+\veps_it_i)}_{+,\Delta_i+\Delta_j,ABC}(z_j,\bar z_j) \nonumber 
\end{align}
\begin{align}
V^{\veps_i}_{+,\Delta_i,AB}(z_i,\bar z_i) \bar V^{\veps_j,CD}_{-,\Delta_j}(z_j,\bar z_j) \sim & \frac{\im\veps_{AB}{}^{CD}\,\veps_i\,\bar z_{ij} }{z_{ij}}\sum_{m=0}^{\infty} \frac{1}{m!}  \int_{\mathbb{R}_+} \frac{\d t_i\,t_i^{\Delta_i-1+m}}{|1+\frac{\veps_i}{\veps_j}t_i|^{\Delta_i+\Delta_j+2}} \\ & \left(\frac{\veps_i\veps_j \bar z_{ij}}{\veps_j+\veps_i\,t_i} \right)^m \bar\partial_j^m\cG^{\mathrm{sgn}(\veps_j+\veps_it_i)}_{-,\Delta_i+\Delta_j}(z_j,\bar z_j) \nonumber \\
&+\frac{\im\veps_{AB}{}^{CD}\,\veps_i\,z_{ji} }{\veps_j\,\bar z_{ji}}\sum_{m=0}^{\infty} \frac{1}{m!}  \int_{\mathbb{R}_+} \frac{\d t_j\,t_j^{\Delta_j-1+m}}{|1+\frac{\veps_j}{\veps_i}t_j|^{\Delta_i+\Delta_j+2}} \nonumber \\ & \left(\frac{\veps_j\veps_i z_{ji}}{\veps_i+\veps_j\,t_j} \right)^m \partial_j^m\cG^{\mathrm{sgn}(\veps_i+\veps_jt_j)}_{+,\Delta_i+\Delta_j}(z_i,\bar z_i) \nonumber \\
\Theta^{\veps_i}_{+,\Delta_i,A}(z_i,\bar z_i) \Xi^{\veps_j}_{+,\Delta_j,BCD}(z_j,\bar z_j) \sim & \frac{
\veps_i\,\veps\,\bar z_{ij} }{z_{ij}}\sum_{m=0}^{\infty} \frac{1}{m!}  \int_{\mathbb{R}_+} \frac{\d t_i\,t_i^{\Delta_i-\frac{3}{2}+m}}{|1+\frac{\veps_i}{\veps_j}t_i|^{\Delta_i+\Delta_j+m}} \\ & \left(\frac{\veps_i \bar z_{ij}}{\veps_j} \right)^m \bar\partial_j^m\Pi^{\mathrm{sgn}(\veps_j+\veps_it_i)}_{\Delta_i+\Delta_j-1,A[BCD]}(z_j,\bar z_j) \nonumber \\
\bar\Theta^{\veps_i,A}_{-,\Delta_i}(z_i,\bar z_i) \Xi^{\veps_j}_{+,\Delta_j,BCD}(z_j,\bar z_j) \sim & \frac{-\im \delta_{[B}{}^A\, z_{ij} }{
\bar z_{ij}}\sum_{m=0}^{\infty} \frac{1}{m!}  \int_{\mathbb{R}_+} \frac{\d t_i\,t_i^{\Delta_i-\frac{3}{2}+m}}{|1+\frac{\veps_i}{\veps_j}t_i|^{\Delta_i+\Delta_j+1}} \\ & \left(\frac{\veps_i\veps_j \, z_{ij}}{\veps_j+\veps_i\,t_i} \right)^m \partial_j^m V^{\mathrm{sgn}(\veps_j+\veps_it_i)}_{+,\Delta_i+\Delta_j,CD]}(z_j,\bar z_j) \nonumber \\
\bar\Theta^{\veps_i,A}_{-,\Delta_i}(z_i,\bar z_i) \Pi^{\veps_j}_{\Delta_j,BCDE}(z_j,\bar z_j) \sim & \frac{-\im \delta_{[B}{}^A\, z_{ij} }{
\bar z_{ij}}\sum_{m=0}^{\infty} \frac{1}{m!}  \int_{\mathbb{R}_+} \frac{\d t_i\,t_i^{\Delta_i-\frac{3}{2}+m}}{|1+\frac{\veps_i}{\veps_j}t_i|^{\Delta_i+\Delta_j+\frac{1}{2}}} \\ & \left(\frac{\veps_i\veps_j\, z_{ij}}{\veps_j+\veps_i\,t_i} \right)^m \partial_j^m\Xi^{\mathrm{sgn}(\veps_j+\veps_it_i)}_{+,\Delta_i+\Delta_j,CDE]}(z_j,\bar z_j) \nonumber \\
\bar\Xi^{\veps_i,ABC}_{-,\Delta_i}(z_i,\bar z_i) \Pi^{\veps_j}_{\Delta_j,DEFG}(z_j,\bar z_j) \sim & \frac{ \veps_{[DEF}{}^{ABC}\, z_{ij} }{\bar z_{ij}}\sum_{m=0}^{\infty} \frac{1}{m!}  \int_{\mathbb{R}_+} \frac{\d t_i\,t_i^{\Delta_i-\frac{1}{2}+m}}{|1+\frac{\veps_i}{\veps_j}t_i|^{\Delta_i+\Delta_j+\frac{3}{2}}} \\ & \left(\frac{\veps_i\veps_j \, z_{ij}}{\veps_j+\veps_i\,t_i} \right)^m \partial_j^m\Theta^{\mathrm{sgn}(\veps_j+\veps_it_i)}_{+,\Delta_i+\Delta_j,G]}(z_j,\bar z_j) \nonumber \\
\Pi^{\veps_i,ABCD}_{\Delta_i}(z_i,\bar z_i) \Pi^{\veps_j}_{\Delta_j,EFGH}(z_j,\bar z_j) \sim & \frac{\im \veps_{EFGH}{}^{ABCD}\,\veps_j\,\bar z_{ij}}{z_{ij}}\sum_{m=0}^{\infty} \frac{1}{m!}  \int_{\mathbb{R}_+} \frac{\d t_i\,t_i^{\Delta_i+m}}{|1+\frac{\veps_i}{\veps_j}t_i|^{\Delta_i+\Delta_j+2}} \label{A.32}\\ & \left(\frac{\veps_i\veps_j \bar z_{ij}}{\veps_j+\veps_i\,t_i} \right)^m \bar\partial_j^m\cG^{\mathrm{sgn}(\veps_j+\veps_it_i)}_{-,\Delta_i+\Delta_j}(z_j,\bar z_j) \nonumber \\
&+ \frac{\im \veps_{EFGH}{}^{ABCD}\,z_{ji}}{\bar z_{ji}}\sum_{m=0}^{\infty} \frac{1}{m!}  \int_{\mathbb{R}_+} \frac{\d t_j\,t_j^{\Delta_j+m}}{|1+\frac{\veps_j}{\veps_i}t_j|^{\Delta_i+\Delta_j+2}} \nonumber \\ 
& \left(\frac{\veps_j\veps_i z_{ji}}{\veps_i+\veps_j\,t_j} \right)^m \partial_i^m\cG^{\mathrm{sgn}(\veps_i+\veps_jt_j)}_{+,\Delta_i+\Delta_j}(z_i,\bar z_i) \nonumber 
\end{align}

\subsection{Supersymmetric Einstein-Yang-Mills}
\begin{align}
    \Theta^{\veps_i}_{+,\Delta_i,A}(z_i,\bar z_i)\,\Gamma^{\msf{a},\veps_j}_{+,\Delta_j,B}(z_j,\bar z_j) \sim & \frac{-\veps_i\veps_j\bar z_{ij}}{z_{ij}}\sum_{m=0}^{\infty} \frac{1}{m!}  \int_{\mathbb{R}_+} \frac{\d t_i\,t_i^{\Delta_i-\frac{3}{2}+m}}{|1+\frac{\veps_i}{\veps_j}t_i|^{\Delta_i+\Delta_j+m}} \\
    &\left(\frac{\veps_i \bar z_{ij}}{\veps_j} \right)^m \bar\partial_j^m\Phi^{\msf{a},\mathrm{sgn}(\veps_j+\veps_it_i)}_{\Delta_i+\Delta_j, AB}(z_j,\bar z_j) \nonumber\\
    \cG^{\veps_i}_{+,\Delta_i}(z_i,\bar z_i)\Gamma^{\msf{a},\veps_j}_{+,\Delta_j,A}(z_j,\bar z_j) \sim & \frac{\veps_i\veps_j\bar z_{ij}}{z_{ij}}\sum_{m=0}^{\infty} \frac{1}{m!}  \int_{\mathbb{R}_+} \frac{\d t_i\,t_i^{\Delta_i-2+m}}{|1+\frac{\veps_i}{\veps_j}t_i|^{\Delta_i+\Delta_j-\frac{1}{2}+m}} \\
    &\left(\frac{\veps_i \bar z_{ij}}{\veps_j} \right)^m \bar\partial_j^m\Gamma^{\msf{a},\mathrm{sgn}(\veps_j+\veps_it_i)}_{+,\Delta_i+\Delta_j, A}(z_j,\bar z_j)\nonumber \\
    \Theta^{\veps_i}_{+,\Delta_i,A}(z_i,\bar z_i)\cO^{\msf{a},\veps_j}_{+,\Delta_j}(z_j,\bar z_j) \sim & \frac{\veps_i\veps_j\bar z_{ij}}{z_{ij}}\sum_{m=0}^{\infty} \frac{1}{m!}  \int_{\mathbb{R}_+} \frac{\d t_i\,t_i^{\Delta_i-\frac{3}{2}+m}}{|1+\frac{\veps_i}{\veps_j}t_i|^{\Delta_i+\Delta_j-\frac{1}{2}+m}} \\
    &\left(\frac{\veps_i \bar z_{ij}}{\veps_j} \right)^m \bar\partial_j^m\Gamma^{\msf{a},\mathrm{sgn}(\veps_j+\veps_it_i)}_{+,\Delta_i+\Delta_j, A}(z_j,\bar z_j) \nonumber\\
    \Theta^{\veps_i}_{+,\Delta_i,A}(z_i,\bar z_i)\,\bar\Gamma^{\msf{a},\veps_j,B}_{-,\Delta_j}(z_j,\bar z_j)\sim & \frac{-\im\veps_i\,\delta_A{}^B\,\bar z_{ij}}{z_{ij}}\sum_{m=0}^{\infty} \frac{1}{m!}  \int_{\mathbb{R}_+} \frac{\d t_i\,t_i^{\Delta_i-\frac{3}{2}+m}}{|1+\frac{\veps_i}{\veps_j}t_i|^{1+\Delta_i+\Delta_j}} \\
    &\left(\frac{\veps_i\veps_j \bar z_{ij}}{\veps_j+\veps_i\,t_i} \right)^m \bar\partial_j^m\cO^{\msf{a},\mathrm{sgn}(\veps_j+\veps_it_i)}_{-,\Delta_i+\Delta_j}(z_j,\bar z_j)\nonumber\\
    \cG^{\veps_i}_{+,\Delta_i}(z_i,\bar z_i)\,\bar\Gamma^{\msf{a},\veps_j,A}_{-,\Delta_j}(z_j,\bar z_j)\sim & \frac{\im\veps_i\,\bar z_{ij}}{z_{ij}}\sum_{m=0}^{\infty} \frac{1}{m!}  \int_{\mathbb{R}_+} \frac{\d t_i\,t_i^{\Delta_i-2+m}}{|1+\frac{\veps_i}{\veps_j}t_i|^{\frac{1}{2}+\Delta_i+\Delta_j}} \\
    &\left(\frac{\veps_i\veps_j \bar z_{ij}}{\veps_j+\veps_i\,t_i} \right)^m \bar\partial_j^m\bar\Gamma^{\msf{a},\mathrm{sgn}(\veps_j+\veps_it_i),A}_{-,\Delta_i+\Delta_j}(z_j,\bar z_j) \nonumber \\
    V^{\veps_i}_{+,\Delta_i,AB}(z_i,\bar z_i)\,\Phi^{\msf{a},\veps_j,CD}_{\Delta_j}(z_j,\bar z_j)\sim & \frac{\im\veps_{AB}{}^{CD} \veps_i\,\bar z_{ij}}{z_{ij}}\sum_{m=0}^{\infty} \frac{1}{m!}  \int_{\mathbb{R}_+} \frac{\d t_i\,t_i^{\Delta_i-1+m}}{|1+\frac{\veps_i}{\veps_j}t_i|^{1+\Delta_i+\Delta_j}}\label{A.15} \\
    &\left(\frac{\veps_i\veps_j \bar z_{ij}}{\veps_j+\veps_i\,t_i} \right)^m \bar\partial_j^m\cO^{\msf{a},\mathrm{sgn}(\veps_j+\veps_it_i)}_{-,\Delta_i+\Delta_j}(z_j,\bar z_j) \nonumber \\
    \cG^{\veps_i}_{+,\Delta_i}(z_i,\bar z_i) \Phi^{\msf{a},\veps_j,AB}_{\Delta_j}(z_j,\bar z_j) \sim & \frac{\im \veps_i\bar z_{ij}}{z_{ij}}\sum_{m=0}^{\infty} \frac{1}{m!}  \int_{\mathbb{R}_+} \frac{\d t_i\,t_i^{\Delta_i-2+m}}{|1+\frac{\veps_i}{\veps_j}t_i|^{\Delta_i+\Delta_j}} \\ & \left(\frac{\veps_i\veps_j \bar z_{ij}}{\veps_j+\veps_i\,t_i} \right)^m \bar\partial_j^m\Phi^{\msf{a},\mathrm{sgn}(\veps_j+\veps_it_i),AB}_{\Delta_i+\Delta_j}(z_j,\bar z_j) \nonumber \\
    \Theta^{\veps_i}_{+,\Delta_i,A}(z_i,\bar z_i) \Phi^{\msf{a},\veps_j,BC}_{\Delta_j}(z_j,\bar z_j) \sim & \frac{-\im \delta_A{}^{[B}\,\veps_i\bar z_{ij} }{z_{ij}}\sum_{m=0}^{\infty} \frac{1}{m!}  \int_{\mathbb{R}_+} \frac{\d t_i\,t_i^{\Delta_i-\frac{3}{2}+m}}{|1+\frac{\veps_i}{\veps_j}t_i|^{\Delta_i+\Delta_j+\frac{1}{2}}} \\ & \left(\frac{\veps_i\veps_j \bar z_{ij}}{\veps_j+\veps_i\,t_i} \right)^m \bar\partial_j^m\bar\Gamma^{\msf{a},\mathrm{sgn}(\veps_j+\veps_it_i),C]}_{-,\Delta_i+\Delta_j}(z_j,\bar z_j) \nonumber 
\end{align}
\newpage
\bibliographystyle{JHEP}
\bibliography{cope111}

\end{document}